\documentclass[preprint,10pt]{elsarticle}

\makeatletter
\def\ps@pprintTitle{%
   \let\@oddhead\@empty
   \let\@evenhead\@empty
   \def\@oddfoot{\reset@font\hfil\thepage\hfil}
   \let\@evenfoot\@oddfoot
}
\makeatother

\usepackage[margin=1in]{geometry}
\usepackage{amsmath,amsfonts,mathtools,amssymb}
\usepackage[T1]{fontenc}\usepackage{lmodern}
\usepackage{booktabs}
\usepackage{subcaption}
\usepackage{xcolor}
\usepackage[hidelinks]{hyperref}
\hypersetup{hidelinks}
\usepackage{multirow}
\usepackage{placeins}
\usepackage{etoolbox}
\usepackage{longtable}
\usepackage{doi}

\definecolor{green}{rgb}{0,0.5,0}

\newcommand{\reviewerOne}[1]{{#1}}
\newcommand{\reviewerTwo}[1]{{#1}}

\newcommand{\sparc}{\texttt{SPARC}}

\begin{document}

\begin{frontmatter}



\title{Code-Verification Techniques for Hypersonic Reacting Flows in Thermochemical Nonequilibrium}


\author[freno]{Brian A.\ Freno}
\ead{bafreno@sandia.gov}
\author[freno]{Brian R.\ Carnes}
\ead{bcarnes@sandia.gov}
\author[freno]{V.\ Gregory Weirs}
\ead{vgweirs@sandia.gov}
\address[freno]{Sandia National Laboratories, PO Box 5800, MS 0828, Albuquerque, NM 87185}

\begin{abstract}
The study of hypersonic flows and their underlying aerothermochemical reactions is particularly important in the design and analysis of vehicles exiting and reentering Earth's atmosphere.  
Computational physics codes can be employed to simulate these phenomena; however, verification of these codes is necessary to certify their credibility.
To date, few approaches have been presented for verifying codes that simulate hypersonic flows, especially flows reacting in thermochemical nonequilibrium.
In this paper, we present our code-verification techniques for \reviewerOne{verifying the spatial accuracy and thermochemical source term in hypersonic reacting flows in thermochemical nonequilibrium.  We demonstrate the effectiveness of these techniques on} the Sandia Parallel Aerodynamics and Reentry Code (\sparc{}).
\end{abstract}


\end{frontmatter}

%


\newcommand{\pz}{\phantom{0}}
\newcommand{\lonenorm}[1]{$L^1$-norm}
\newcommand{\ltwonorm}[1]{$L^2$-norm}
\newcommand{\linfnorm}[1]{$L^\infty$-norm}

\newcommand{\globalaccuracy}{p}
\newcommand{\rhovec}{\boldsymbol{\rho}}
\newcommand{\vel}{\mathbf{v}}
\newcommand{\xvel}{u}
\newcommand{\yvel}{v}
\newcommand{\zvel}{w}
\newcommand{\temperature}{T}

\newcommand{\state}{\mathbf{U}}
\newcommand{\fluxc}{\mathbf{F}_c}
\newcommand{\fluxd}{\mathbf{F}_d}
\newcommand{\fluxp}{\mathbf{F}_p}
\newcommand{\source}{\mathbf{S}}
\newcommand{\sdflux}{\mathbf{J}}
\newcommand{\vstress}{\boldsymbol{\tau}}
\newcommand{\nspecies}{{n_s}}
\newcommand{\Runiv}{\bar{R}}
\newcommand{\Ntwo}{\text{N}_2}
\newcommand{\Otwo}{\text{O}_2}
\newcommand{\wvecdot}{\dot{\mathbf{w}}}
\newcommand{\wdot}{\dot{w}}
\newcommand{\nvs}{{n_{v_s}}}
\newcommand{\Qtv}{Q_{t-v}}
\newcommand{\evecv}{\mathbf{e}_v}
\newcommand{\ev}{e_v}
\newcommand{\evs}[1]{e_{v_{#1}}}
\newcommand{\evsm}{\evs{s,m}}
\newcommand{\thetavsm}{\theta_{v_{s,m}}}

\newcommand{\mach}{M}
\newcommand{\soundspeed}{a}
\newcommand{\scalar}{\alpha}
\newcommand{\errorvec}{\mathbf{e}}
\newcommand{\errorscalar}{e}
\newcommand{\ncells}{n}

\newcommand{\rgas}{R}
\newcommand{\residual}{\mathbf{r}}

\newcommand{\veld}{\mathbf{u}}

\newcommand{\nreact}{{n_r}}

\newcommand{\statedummy}{\mathbf{V}}

\newcommand{\cellcentroid}{\mathbf{x}}
\newcommand{\vol}{\Omega}
\newcommand{\errornorm}{\varepsilon}
\newcommand{\perturbation}{\epsilon}
\newcommand{\parameters}{\boldsymbol{\mu}}

\section{Introduction} 

Hypersonic flows are distinguished by flow velocities and stagnation enthalpies that are high enough to induce chemical reactions and excitation of thermal energy modes.  In particular, the time scales of the reactions and thermal excitation are comparable to the characteristic flow time, requiring a fully coupled modeling approach. 
The study of hypersonic flows and their underlying aerothermochemical phenomena is particularly important in the design and analysis of vehicles exiting and reentering Earth's atmosphere~\cite{gnoffo_1989,anderson_2006}.  
For much of this flight regime, the gas can be modeled as a continuum.
The relevant chemical species are each tracked, such that the interactions within the flow field are accounted for in the evolution of the chemical state.  Similarly, considering each internal energy mode allows the different modes to relax towards equilibrium while accounting for the flow and chemical states.  
As the flow velocity or stagnation enthalpy increases, more complex chemical kinetics and internal energy models must be included, and ionization, radiation, and other phenomena may become important.  

Verification and validation of the implementation and suitability of such models are necessary to develop confidence in the credibility of the simulations.
Validation assesses how well the models instantiated in the code represent the relevant physical phenomena.  It is typically performed by comparing simulation results with experimental results to assess the suitability of the models, the model error, and the practical bounds of validity of the models.
On the other hand, verification assesses the accuracy of the numerical solutions the code produces, relative to the assumptions and expectations associated with the numerical methods.  Following Roache~\cite{roache_1998},  Salari and Knupp~\cite{salari_2000}, and Oberkampf and Roy~\cite{oberkampf_2010}, verification can be divided into code verification and solution verification.  Solution verification focuses on numerical-error estimation for a particular simulation, whereas code verification focuses on the correctness of the numerical-method implementation in the code.  A review of code and solution verification is presented by Roy~\cite{roy_2005}.

This paper focuses on code verification.  \reviewerOne{When numerically solving the underlying equations of the aforementioned models}, the equations must be discretized using, for example, finite differences, finite volumes, or finite elements.  Due to the finite nature of the discretization, the equations incur a truncation error and, consequently, their solutions introduce a discretization error.  As the discretization is refined, the discretization error should decrease.  More rigorously, a code should achieve an expected order of accuracy: as the mesh is refined by a factor, the error should decrease at a rate that is an expected power of that factor, provided the mesh is in the asymptotic region.  In practice, since the exact solution is generally unavailable, manufactured solutions are frequently employed~\cite{roache_2001}.

Code verification has been performed on computational physics codes associated with several physics disciplines, including fluid dynamics~\cite{roy_2004,bond_2007,veluri_2010,oliver_2012}, solid mechanics~\cite{chamberland_2010}, fluid--structure interaction~\cite{etienne_2012}, heat transfer in fluid--solid interaction~\cite{veeraragavan_2016}, multiphase flows~\cite{brady_2012}, radiation hydrodynamics~\cite{mcclarren_2008}, electrodynamics~\cite{ellis_2009}, and electromagnetism~\cite{marchand_2013}.  
Though not as common, code-verification techniques for hypersonic flows have been presented by Roy et al.~\cite{roy_2003} for a single-species perfect gas and by Gollan and Jacobs~\cite{gollan_2013} for a multi-species gas in thermal equilibrium.

In this paper, we discuss the code-verification techniques we have employed for flows without discontinuities.  The most noteworthy contribution of this work is our approach to verifying hypersonic reacting flows in thermochemical nonequilibrium.  
\reviewerOne{As the scope of this paper is limited to code verification, subsequent instances of `verification' are used to abbreviate `code verification'.

To assess the spatial discretization, we employ manufactured and exact solutions on uniform and nonuniform meshes and study the convergence of the spatial error norms.  Because we consider smooth flows, we employ the more rigorous \linfnorm{}, of which we demonstrate the greater effectiveness in detecting deviations in the expected order of accuracy.  These tests identified lower-order boundary-condition implementations, which reduced the convergence rate and spatial accuracy before being corrected.

However, while these techniques are effective for assessing the spatial discretization, they do not directly reveal errors in the algebraic thermochemical source term.  Therefore, to verify these terms, we compare with an independently developed code for thousands of samples, spanning extreme conditions, with the expectation that the values agree to machine precision.}  \reviewerTwo{Additionally, we examine the convergence of the statistics with respect to the number of samples to assess their sufficiency.}  \reviewerOne{This work exposed disagreement due to convergence criteria, as well as errors in model parameters, resulting in corrections to both.}

This paper is organized as follows. 
Section~\ref{sec:governing_equations} describes the governing conservation, energy-exchange, and chemical-kinetics equations.  \reviewerOne{%
Section~\ref{sec:spatial_accuracy} details our approach for verifying the spatial discretization through the use of manufactured and exact solutions.  
Section~\ref{sec:spatial_results} demonstrates the effectiveness of the verification techniques for the spatial discretization. 
Section~\ref{sec:source_terms} presents our approach for verifying the thermochemical source term.}
Section~\ref{sec:rg_results} demonstrates the effectiveness of the verification techniques for the thermochemical source term.
Section~\ref{sec:conclusions} provides conclusions and an outlook for future work.

\section{Governing Equations} 
\label{sec:governing_equations}

In this paper, we consider hypersonic reacting flows in thermochemical nonequilibrium.  We make the following approximations: (1) electronic energy is negligible, (2) vibrationally excited molecules can be characterized by a single vibrational temperature $T_v$, and (3) the translational and rotational temperatures are in thermal equilibrium: $T_t=T_r=T$.

\subsection{Conservation Equations}
With these approximations, we model the conservation of mass, momentum, and energy for a gas in thermochemical nonequilibrium~\cite{gnoffo_1989,anderson_2006}:
\begin{align}
\frac{\partial \state}{\partial t}%
+ %
\nabla\cdot \mathbf{F}_c\left(\state\right) %
&= %
- %
\nabla\cdot \mathbf{F}_p\left(\state\right) %
+ %
\nabla\cdot \mathbf{F}_d\left(\state\right) %
+ %
\mathbf{S}\left(\state\right),
\label{eq:conservation}
\end{align}
where the conservative-variable state vector $\state$, convective flux $\fluxc$, pressure flux $\fluxp$, diffusive flux $\fluxd$, and thermochemical source term $\source$ are
\begin{gather*}
\allowdisplaybreaks
\state %
= %
\left\{%
\begin{matrix}
\rhovec \\
\rho\vel \\
\rho E \\
\rho \ev
\end{matrix}
\right\},
\quad
\fluxc\left(\state\right) %
= %
\left[%
\begin{matrix}
\rhovec\vel^T \\
\rho\vel\vel^T \\
\rho E \vel^T \\
\rho \ev \vel^T
\end{matrix}
\right],
\quad
\fluxp\left(\state\right) %
= %
\left[
\begin{matrix}
\mathbf{0} \\
p\mathbf{I} \\
p \vel^T \\
\phantom{^T}\mathbf{0}^T
\end{matrix}
\right],
\quad
\fluxd\left(\state\right) %
= %
\left[
\begin{matrix}
-\sdflux \\
\vstress \\
\left(\vstress \vel - \mathbf{q}  - \mathbf{q}_v - \sdflux^T\mathbf{h}\right)^T \\
\left(- \mathbf{q}_v- \sdflux^T\evecv\right)^T
\end{matrix}
\right],
\nonumber
\\[0.5em] 
\source\left(\state\right) %
= %
\left\{
\begin{matrix}
\wvecdot \\
\mathbf{0} \\
0 \\
\Qtv+\evecv^T\wvecdot
\end{matrix}
\right\}.
\end{gather*}
%
\reviewerOne{$\rhovec=\left\{\rho_1,\hdots,\rho_\nspecies\right\}^T$ is the vector of the densities of the $\nspecies$ species, and
$\rho=\sum_{s=1}^{\nspecies}\rho_s$ is the mixture density.
For the particular case of air with $\nspecies=5$, which is considered later in this paper, the species are $\Ntwo$, $\Otwo$, NO, N, and O.}
$\vel=\left\{\xvel,\,\yvel,\,\zvel\right\}^T$ is the velocity vector, 
$p=\sum_{s=1}^{\nspecies}\frac{\rho_s}{M_s}\Runiv T$ is the pressure, \reviewerOne{$M_s$ is the molecular weight of species $s$, $\Runiv =8314.47\text{ J}/\text{kmol}/\text{K}$ is the universal gas constant, and $T$ is the translational--rotational temperature.
$\sdflux= \left[\rho_1\mathbf{u}_1,\hdots,\rho_\nspecies\mathbf{u}_\nspecies\right]^T$ is the diffusion flux of the species, and $\veld_s$ is the diffusion velocity of species $s$.}
$\vstress$ is the viscous stress tensor,
$\mathbf{q}$ is the heat flux,
$\mathbf{q}_v$ is the vibrational heat flux, and
$\mathbf{h}=\left\{h_1,\hdots,h_\nspecies\right\}^T$ is the vector of enthalpies per mass of the species.
\reviewerOne{$\wvecdot=\left\{\wdot_1,\hdots,\wdot_\nspecies\right\}^T$ is the vector of mass production rates of the species per volume, and $\Qtv$ is the translational--vibrational energy exchange.}

At high temperatures, vibrational energy is internally excited within molecules, which can cause dissociation or ionization within the gas~\cite{anderson_2006}.
The mixture vibrational energy per mass is 
\begin{align*}
\ev=\sum_{s=1}^{\nspecies}\frac{\rho_s}{\rho} \evs{s},
\end{align*}
where
$\evecv=\left\{\evs{1},\hdots,\evs{\nspecies}\right\}^T$ is the vector of vibrational energies per mass of the species, such that
\begin{align*}
\evs{s} =%
\left\{\begin{array}{@{}c l@{}}
\sum_{m=1}^\nvs\evsm(T_v) & \text{for molecules,} \\[0.5em] 
0 & \text{for atoms.}\end{array}\right.
\end{align*}
The vibrational energy per mass of mode $m$ of species $s$ is assumed to have a Boltzmann distribution over the vibrational energy levels~\cite{gnoffo_1989,anderson_2006}:
\begin{align*}
\evsm(T')= %
\frac{\Runiv }{M_s}\frac{\thetavsm}{\exp\left(\thetavsm/T'\right)-1},
\end{align*}
\reviewerOne{where} $\nvs$ is the number of vibrational modes of species $s$ ($\nvs=0$ for atoms), and $\thetavsm$ is the characteristic vibrational temperature of mode $m$ of species $s$.  \reviewerOne{$\thetavsm$ is typically obtained from experiments, and we retrieve the value from a lookup table, such as Table~\ref{tab:theta_v}.}

The total energy per mixture mass is
\begin{align*}
E= \frac{|\vel|^2}{2} + \sum_{s=1}^\nspecies \frac{\rho_s}{\rho}\left(c_{\mathcal{V}_s}T+\evs{s}+h_s^o\right),
\end{align*}
\reviewerOne{where} $c_{\mathcal{V}_s}$ is the specific heat at constant volume of species $s$, which is the sum of the translational and rotational contributions.  For \reviewerOne{diatomic} molecules, $c_{\mathcal{V}_s}=\frac{5}{2}\frac{\Runiv }{M_s}$ and, for atoms, $c_{\mathcal{V}_s}=\frac{3}{2}\frac{\Runiv }{M_s}$.  $h_s^o$ is the formation enthalpy of species $s$.

\subsection{Diffusive-Term Modeling}
We use Fick's first law to model the diffusion flux of the species : $\sdflux=-\rho \mathbf{D}\nabla\frac{\rhovec}{\rho}$, where $\mathbf{D}=\text{diag}\left\{D_1,\hdots,D_\nspecies\right\}$ contains the effective diffusion coefficients of the species.
Assuming the flow is Newtonian and satisfies Stokes' hypothesis, we model the viscous stress tensor as $\vstress=\mu\left(\nabla\vel+\left(\nabla\vel\right)^T\right)-\frac{2}{3}\mu\left(\nabla\cdot\vel\right)\mathbf{I}$, where $\mu$ is the dynamic viscosity.  
To model the heat fluxes, we employ Fourier's law: $\mathbf{q}=-\kappa\nabla T$ and $\mathbf{q}_v=-\kappa_v\nabla T_v$, where $\kappa$ and $\kappa_v$ are, respectively, the translational--rotational and vibrational heat conductivities of the mixture.  
$\mu$ and $\kappa$ for the mixture are computed using \reviewerOne{Wilke's mixing rule~\cite{wilke_1950}}.

\subsection{Translational--Vibrational Energy Exchange Modeling}
The translational--vibrational energy exchange is computed using the Landau--Teller model~\cite{landau_1936}: 
\begin{align}
\Qtv=\sum_{s=1}^{n_s}\rho_s\sum_{m=1}^\nvs \frac{\evsm(T)-\evsm(T_v)}{\left\langle\tau_{s,m}\right\rangle},
\label{eq:landau_teller}
\end{align}
where 
$\left\langle\tau_{s,m}\right\rangle$ is the translational--vibrational energy relaxation time for mode $m$ of species $s$.  

To compute the relaxation time, 
we use the semi-empirical approach of Millikan and White~\cite{millikan_1963}, which is extended to account for higher temperatures~\cite{park_1993,park_1994}:
\begin{align*}
\left\langle\tau_{s,m}\right\rangle = \left(\sum_{s'=1}^\nspecies \frac{y_{s'}}{\tau_{s,m,s'}}\right)^{-1} %
+ %
\left[\left(N_\text{A}\sum_{s'=1}^{n_s}\frac{\rho_{s'}}{M_{s'}}\right)\sigma_{v_s}\sqrt{\frac{8}{\pi}\frac{\Runiv T}{M_s}}\right]^{-1},
\end{align*}
where \reviewerOne{the mole fraction of species $s$ is}
\begin{align*}
y_s = \frac{\rho_s/M_s}{\sum_{s'=1}^\nspecies\rho_{s'}/M_{s'}},
\end{align*}
$N_\text{A}=6.022140857\times 10^{26}\text{ kmol}^{-1}$ is the Avogadro constant, and \reviewerOne{the translational--vibrational energy relaxation time for mode $m$ of species $s$ when colliding with species $s'$ is modeled by}
\begin{align}
\tau_{s,m,s'} = \frac{\exp\left[a_{s,m,s'}\left(T^{-1/3}-b_{s,m,s'}\right)-18.42\right]}{p'}.
\label{eq:tau}
\end{align}
$p'$ is the pressure expressed in atmospheres.  For many gases, $a_{s,m,s'}$ and $b_{s,m,s'}$ can be modeled by
\begin{align}
a_{s,m,s'}&{}=1.16\times 10^{-3}\mu_{s,s'}^{1/2}\thetavsm^{4/3}, \nonumber
\\
b_{s,m,s'}&{}=0.015\mu_{s,s'}^{1/4},
\label{eq:ab}
\end{align}
where $\mu_{s,s'}=\frac{M_s M_{s'}}{M_s + M_{s'}}$ is the reduced mass of species $s$ and $s'$.
Additionally, \reviewerOne{the collision-limiting vibrational cross section is modeled by}
\begin{align*}
\sigma_{v_s} = \sigma_{v_s}' \left(\frac{\text{50,000 K}}{T}\right)^2,
\end{align*}
where $\sigma_{v_s}'$ is the collision-limiting vibrational cross section at 50,000 K.
\subsection{Chemical-Kinetics Modeling}
The mass production rate per volume for species $s$ is modeled by
\begin{align}
\wdot_s = M_s \sum_{r=1}^\nreact \left(\beta_{s,r}-\alpha_{s,r}\right)\left(R_{f_r}-R_{b_r}\right),
\label{eq:ckm}
\end{align}
where $\alpha_{s,r}$ and $\beta_{s,r}$ are, respectively, the reactant and product stoichiometric coefficients for species $s$ in reaction $r$.  $R_{f_r}$ and $R_{b_r}$ are the forward and backward reaction rates for reaction $r$.

The reaction rates are defined by
\begin{align}
R_{f_r} &{}= \gamma k_{f_r}\prod_{s=1}^\nspecies \left(\frac{1}{\gamma}\frac{\rho_s}{M_s}\right)^{\alpha_{s,r}},
\label{eq:forward_rate}
\\
R_{b_r} &{}= \gamma k_{b_r}\prod_{s=1}^\nspecies \left(\frac{1}{\gamma}\frac{\rho_s}{M_s}\right)^{ \beta_{s,r}},
\label{eq:backward_rate}
\end{align}
such that, in \eqref{eq:forward_rate} and~\eqref{eq:backward_rate}, $k_{f_r}$, $k_{b_r}$, and $\rho_s/M_s$ are expressed using the centimeter--gram--second system of units (CGS), $R_{f_r}$ and $R_{b_r}$ are expressed using the meter--kilogram--second system of units, and $\gamma= 1000 \text{ }\frac{\text{cm}^3\cdot\text{kmol}}{\text{m}^3\cdot\text{mol}}$ is the conversion factor.
The forward and backward reaction-rate coefficients $k_{f_r}$ and $k_{b_r}$ are modeled using the approach of Park~\cite{park_1990}:
\begin{align*}
k_{f_r}(T_c) &{}= C_{f_r} T_c^{\eta_r}\exp\left(-\theta_r/T_c\right),
\\
k_{b_r}(T) &{}= \frac{k_{f_r}(T)}{K_{e_r}(T)},
\end{align*}
where 
$C_{f_r}$ and $\eta_r$ are empirical parameters, and
$\theta_r$ is the activation energy of reaction $r$, divided by the Boltzmann constant.
$T_c$ is the rate-controlling temperature; it is set to $T_c=\sqrt{T T_v}$ for dissociative reactions and $T_c=T$ for exchange reactions.
$K_{e_r}$ is the equilibrium constant for reaction $r$, modeled by
\begin{align}
K_{e_r}(T) = \exp\left[A_{1_r} \left(\frac{T}{\text{10,000 K}}\right) + A_{2_r} + A_{3_r} \ln \left(\frac{\text{10,000 K}}{T}\right) + A_{4_r} \frac{\text{10,000 K}}{T} + A_{5_r} \left(\frac{\text{10,000 K}}{T}\right)^2\right],
\label{eq:keq}
\end{align}
where $A_{i_r}$ are empirical curve-fit coefficients.

In this paper, $K_{e_r}(T)$ is limited to $[\exp(-81),\,\exp(81)]$, and, when computing $\wvecdot$, $T$ and $T_v$ are increased to 500 K if they are less than 500 K.

\subsection{Gas Modeling} 
\label{subsec:five_spec}

We limit the scope of this paper to the five-species air model, which consists of $\Ntwo$, $\Otwo$, NO, N, and O.  These species can undergo the dissociation and exchange reactions listed in Table~\ref{tab:reactions}.  Additional properties of the species and their reactions are provided in \ref{appx}.

\begin{table}[htbp!]
\centering
\begin{tabular}{c r @{${}+{}$} c @{${}\leftrightharpoons{}$} l @{${}+{}$} c @{${}+{}$} l c c}
\toprule
$r$ & \multicolumn{5}{c}{Reaction}                              &  & Type of Reaction \\
\midrule
\pz1--5\pz& $\Ntwo    $ & $\mathcal{M}      $ & $\text{N}$ & $\text{N}$ & $\mathcal{M},   $ & $\mathcal{M}=\{\Ntwo,\,\Otwo,\,\text{NO},\,\text{N},\,\text{O}\}$ &  Dissociation \\
\pz6--10  & $\Otwo    $ & $\mathcal{M}      $ & $\text{O}$ & $\text{O}$ & $\mathcal{M},   $ & $\mathcal{M}=\{\Ntwo,\,\Otwo,\,\text{NO},\,\text{N},\,\text{O}\}$ &  Dissociation \\
11--15    & $\text{NO}$ & $\mathcal{M}      $ & $\text{N}$ & $\text{O}$ & $\mathcal{M},   $ & $\mathcal{M}=\{\Ntwo,\,\Otwo,\,\text{NO},\,\text{N},\,\text{O}\}$ &  Dissociation \\
  16      & $\Ntwo    $ & $\text{O}         $ & $\text{N}$ & \multicolumn{2}{@{}l}{$\text{NO}$} &                                                                 &  Exchange \\
  17      & $\text{NO}$ & $\text{O}         $ & $\text{N}$ & \multicolumn{2}{@{}l}{$\Otwo$}     &                                                                 &  Exchange \\
\bottomrule
\end{tabular}
\caption{Five-species air model: Reactions.}
\label{tab:reactions}
\end{table}

\reviewerOne{We note that the five-species air model is just one of a few options.  Gimelshein et al.~\cite{gimelshein_2019,gimelshein_2020} provide a comparison of air models represented by different numbers of species.  Nonetheless, the verification techniques we propose for the thermochemical source term are applicable to any number of species.} 

\FloatBarrier

\section{Verification Techniques for Spatial Discretization} 
\label{sec:spatial_accuracy}

We begin our approach to code verification by computing the spatial accuracy of the numerical discretization.
To compute the spatial accuracy, we compare the solution to the discretized equations with the solution to the continuous equations.  For each of the flow variables, we compute error norms and compare the rates at which the error norms decrease with respect to the rates at which the mesh size increases.

\subsection{Spatial Accuracy} 
In a steady state, the governing system of partial differential equations~\eqref{eq:conservation} can be written generally as
\begin{align}
\residual(\state)=\mathbf{0},
\label{eq:orig}
\end{align}
where $\residual$ is the residual vector, and $\state=\state(\cellcentroid)$ is the state vector.
To solve~\eqref{eq:orig} numerically, it must be discretized:
\begin{align}
\residual_h(\state_h)=\mathbf{0},
\label{eq:disc}
\end{align}
where $\residual_h$ is the residual of the discretized system of equations, $\state_h$ is the solution to the discretized equations, \reviewerOne{and $h$ describes the size of the mesh employed by the discretization}.
To simplify the notation in this section, we assume appropriate mappings of the residuals and solutions onto continuous or discrete space.

The truncation error is 
\begin{align}
\boldsymbol{\tau}_h(\statedummy)%
=%
\residual_h(\statedummy) - \residual(\statedummy).
\label{eq:trunc1}
\end{align}
Letting $\statedummy=\state_h$ and adding \eqref{eq:orig}, \eqref{eq:trunc1} becomes
\begin{align*}
\boldsymbol{\tau}_h(\state_h)%
=%
\residual_h(\state_h) - \residual(\state_h) + \residual(\state)%
=%
\residual(\state) - \residual(\state_h)
.
\end{align*}
When $\residual$ is linear or linearized with respect to $\state$, the discretization error $\errorvec_h = \state_h - \state$ is related to the truncation error by $\residual(\errorvec_h)=-\boldsymbol{\tau}_h(\state_h)$~\cite{ferziger_2002,oberkampf_2010}.

For a $\globalaccuracy^\text{th}$-order-accurate discretization, the truncation error is
\begin{align*}
\boldsymbol{\tau}_h%
=%
\mathbf{C}_\residual h^{\globalaccuracy} + \mathcal{O}(h^{\globalaccuracy+1}),
\end{align*}
where $\mathbf{C}_\residual$ is a function of derivatives of the state vector but is independent of $h$. 
Once the meshes are fine enough that the approximation is within the asymptotic region, $h^{\globalaccuracy+1}\ll h^{\globalaccuracy}$, then 
$\boldsymbol{\tau}_h\approx\mathbf{C}_\residual h^{\globalaccuracy}$ and
$\errorvec_h\approx\mathbf{C}_\state h^{\globalaccuracy}$, where $\mathbf{C}_\state$ is also independent of $h$.

The observed order of accuracy can be computed using two meshes.  For example, let the coarser mesh be characterized by $h$.  The second mesh, if $q$-times as fine in each dimension as the first, is characterized by $h/q$.  For each scalar field $\scalar$ in $\state$ (e.g., $\scalar=\{\rho_1,\hdots,\rho_{n_s},\,\xvel,\,\yvel,\,\zvel,\,T,\,T_v\}$), each mesh has a discretization error: $\errorscalar_h^\scalar\approx C_\scalar h^{\globalaccuracy}$ and $\errorscalar_{h/q}^\scalar\approx C_\scalar (h/q)^{\globalaccuracy}$.  $\globalaccuracy$ can be approximated locally by
\begin{align}
\globalaccuracy\approx\frac{\log \big|\errorscalar_h^\scalar/\errorscalar_{h/q}^\scalar\big|}{\log q}=\log_q \big|\errorscalar_h^\scalar/\errorscalar_{h/q}^\scalar\big|. 
\label{eq:accuracy}
\end{align}

\subsection{Solutions} 
Computing the order of accuracy $\globalaccuracy$ in~\eqref{eq:accuracy} requires computing errors $\errorvec_h$, which, in turn, require solutions $\state$.  Therefore, we employ exact and manufactured solutions.

\subsubsection{Exact Solutions}
For limited cases, there exist exact solutions $\state_\text{Exact}$ to~\eqref{eq:orig}, which can be directly compared with the computed solutions $\state_h$ from~\eqref{eq:disc} with negligible implementation effort.  However, available exact solutions only span a small subset of the application space, so we require additional approaches to thoroughly test the capabilities of the code.

\subsubsection{Manufactured Solutions}
Manufactured solutions $\state_\text{MS}$ enable us to develop solutions that exercise the features we intend to test.  Unless the manufactured solutions are exact solutions, they will not satisfy~\eqref{eq:orig}: $\residual(\state_\text{MS})\ne\mathbf{0}$.  Therefore, a forcing term is added to~\eqref{eq:disc} to account for the presence of the manufactured solutions:
\begin{align}
\residual_h(\state_h)=\residual(\state_\text{MS}).
\label{eq:mms}
\end{align}
$\residual(\state_\text{MS})$ in~\eqref{eq:mms} is computed analytically since $\residual$ and $\state_\text{MS}$ are known.  

Because the equations are differential and the error is a function of derivatives of the state vector, the manufactured solutions should be smooth, continuously differentiable functions with generally nonzero derivatives.  Additionally, for the approximation to be in the asymptotic region without requiring especially fine meshes, variations over the domain should not be large.

\subsection{Error Norms} 
The order of accuracy $\globalaccuracy$ can be computed at a single location in the domain, as is done in~\eqref{eq:accuracy}; however, this approach has two shortcomings: (1) for cell-centered schemes, the cell centroids of a coarser mesh only coincide with those arising after mesh refinement in very limited cases, and (2) in regions where the errors vanish, the computed order of accuracy is meaningless.  Therefore, for each scalar field $\scalar$, we use error norms $\errornorm_{\scalar}$ to quantify the order of accuracy $\globalaccuracy$:
\begin{align}
\globalaccuracy=\log_q \left(\errornorm_{\scalar_h}/\errornorm_{\scalar_{h/q}}\right). 
\label{eq:accuracy_norm}
\end{align}
Because the error norms are global quantities, they do not require coincident solution locations across meshes, and they will only be zero for trivial flows.

We consider two norms in particular:
\begin{enumerate}
\item \lonenorm{}: 
\begin{align}
\errornorm_\scalar^1 %
= %
\|\scalar_h(\cellcentroid)-\scalar(\cellcentroid)\|_1 %
= %
\int_\vol |\scalar_h(\cellcentroid)-\scalar(\cellcentroid)| d\vol.
\label{eq:l1norm}
\end{align}

The \lonenorm{} enables us to compute the spatial order of accuracy based on the average error throughout the domain without significant contamination from localized deviations.  Such localized deviations can arise from discontinuities, such as shocks, as well as from boundary conditions discretized with lower-order spatial accuracy than what is used for the domain interior.

\item \linfnorm{}: 
\begin{align}
\errornorm_\scalar^\infty=\|\scalar_h(\cellcentroid)-\scalar(\cellcentroid)\|_\infty=\max_{\mathbf{x}\in\vol}\left|\scalar_h(\cellcentroid)-\scalar(\cellcentroid)\right|. 
\label{eq:linfnorm}
\end{align}

Unlike the \lonenorm{}, the \linfnorm{} catches the aforementioned localized deviations by computing the maximum error throughout the domain.  This is particularly useful when such deviations are unexpected.
\end{enumerate}

For the smooth flows considered in this paper, the observed orders of accuracy computed by the \lonenorm{} and \linfnorm{} are expected to be the same.
\section{Spatial-Discretization Verification Results} 
\label{sec:spatial_results}


\reviewerOne{We demonstrate the aforementioned and forthcoming code-verification techniques on the Sandia Parallel Aerodynamics and Reentry Code} (\sparc{})~\cite{howard_2017,carnes_2019,kieweg_2019,ray_2019} presently being developed at Sandia National Laboratories.  \sparc{} is a compressible computational fluid dynamics code designed to model transonic and hypersonic reacting turbulent flows.

\sparc{} employs a cell-centered finite-volume discretization, and the simulations presented herein use the Steger--Warming flux-vector splitting scheme.
The expectation is that \sparc{} is second-order accurate $(\globalaccuracy=2)$ \reviewerOne{in space} for flows without discontinuities.  Therefore, because the midpoint rule has a discretization error of $\mathcal{O}(h^2)$, we integrate $\residual(\state_\text{MS})$ in \eqref{eq:mms} using the midpoint rule, without decreasing the order of accuracy. 
We additionally use approximations to compute the \lonenorm{} in \eqref{eq:l1norm} and the \linfnorm{} in \eqref{eq:linfnorm}:
\begin{align}
\errornorm_\scalar^1 %
&{}\approx %
\sum_{i=1}^{\ncells}\left(\vol_i\left|\scalar_h(\mathbf{x}_i)-\scalar(\mathbf{x}_i)\right|\right),
\label{eq:l1_approx}
\\[.5em]
\errornorm_\scalar^\infty %
&{}\approx %
\max_{1\le i\le \ncells}\left|\scalar_h(\mathbf{x}_i)-\scalar(\mathbf{x}_i)\right|,
\label{eq:linf_approx}
\end{align}
where $\mathbf{x}_i$ and $\vol_i$ are, respectively, the centroid and volume of cell $i$, \reviewerOne{and $\ncells$ is the number of cells}.  For a uniform mesh, \eqref{eq:l1_approx} is the same as the discrete \lonenorm{}, when multiplied by the cell volume $\vol_i$.  Equation \eqref{eq:linf_approx} is the discrete \linfnorm{}.

\sparc{} computes $\residual(\state_\text{MS})$ in~\eqref{eq:mms} through the automatic differentiation tool Sacado~\cite{phipps_2012,bartlett_2006}.  This avoids the need for using an external package to compute $\residual(\state_\text{MS})$ through symbolic manipulation.  
A major drawback to using symbolic manipulation is the large amount of output generated, which must be copied and formatted into a source-code file.  For the work in this paper, a single term for one equation would have required several lines of code.  On the other hand, the automatic-differentiation approach we employ only requires coding the terms in the differential equation.  By templating the exact solution, we can generate the equivalent source term using only a few lines of code.  The source term is never provided in symbolic form, but it can be efficiently evaluated.  The Sacado implementation in \sparc{} has been unit tested by comparing with examples computed using symbolic manipulation.  The results of those tests agree within machine precision.

For these results, the goal is to ensure second-order accuracy is achievable.  
Therefore, limiters are disabled and boundary conditions are expected to be second-order accurate.  
For complex simulations, the accuracy may be deliberately reduced in favor of numerical stability.  The implications of which can be assessed through solution verification, which is not the focus of this paper.

Our verification of \sparc{} begins with the simplest tests, and, as these tests are satisfied, we add complexity to the subsequent tests.  We begin by verifying single-species supersonic inviscid flow, ultimately increasing the complexity to multi-species hypersonic inviscid flow in thermochemical nonequilibrium.

\begin{figure}[h!]
\centering
\includegraphics[scale=.22,clip=true,trim=0in 0in 0in 0in]{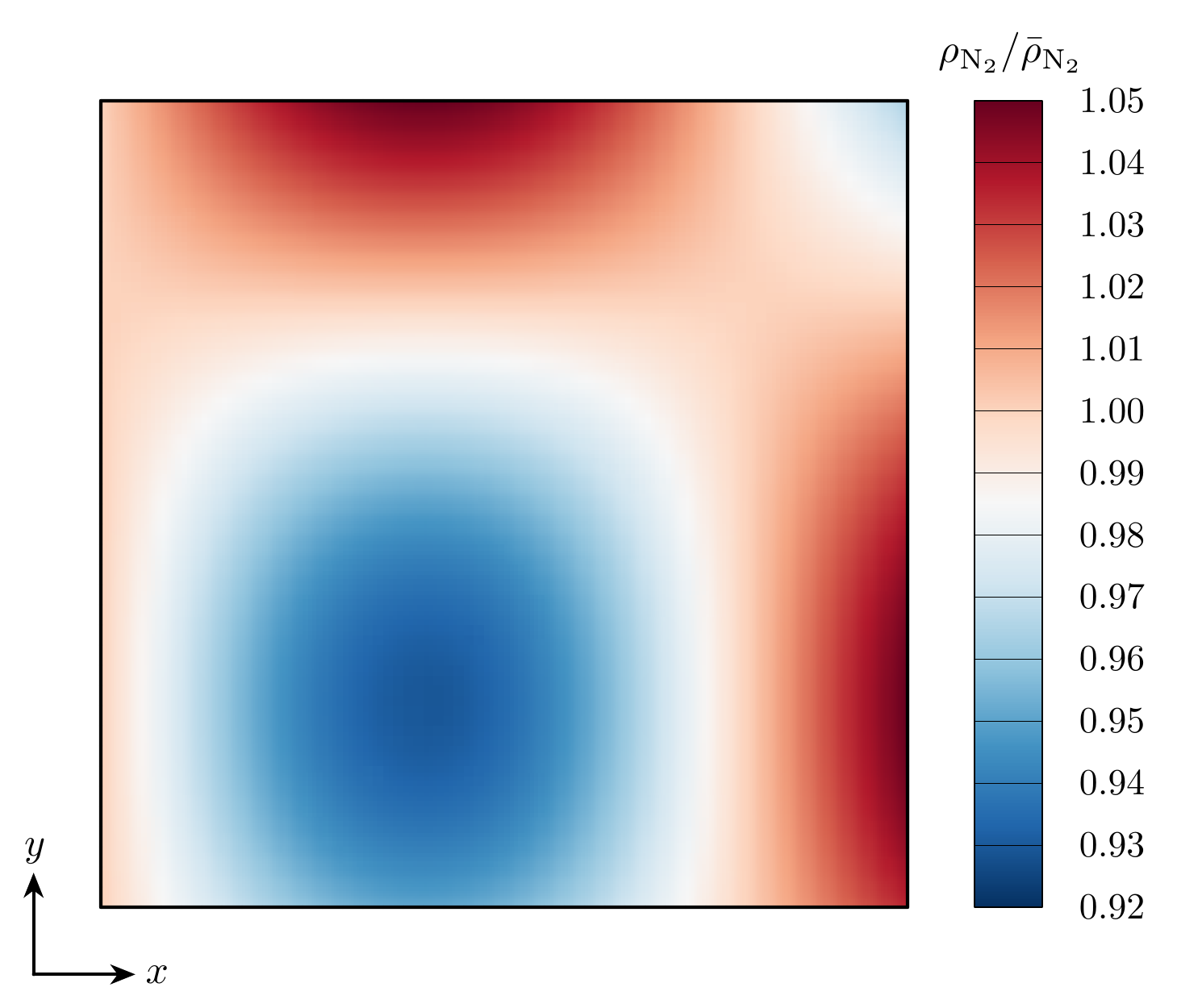}
\includegraphics[scale=.22,clip=true,trim=0in 0in 0in 0in]{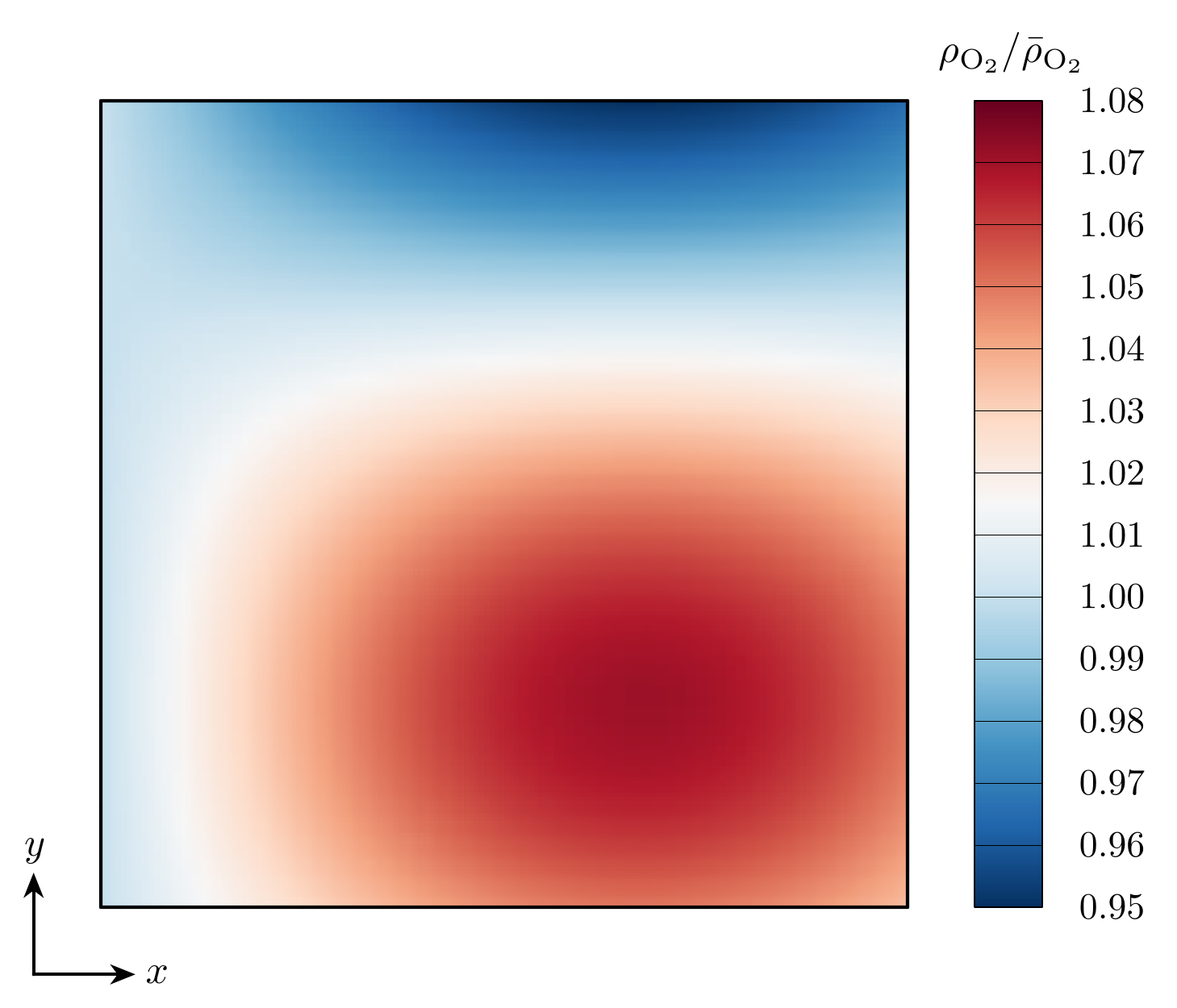}
\includegraphics[scale=.22,clip=true,trim=0in 0in 0in 0in]{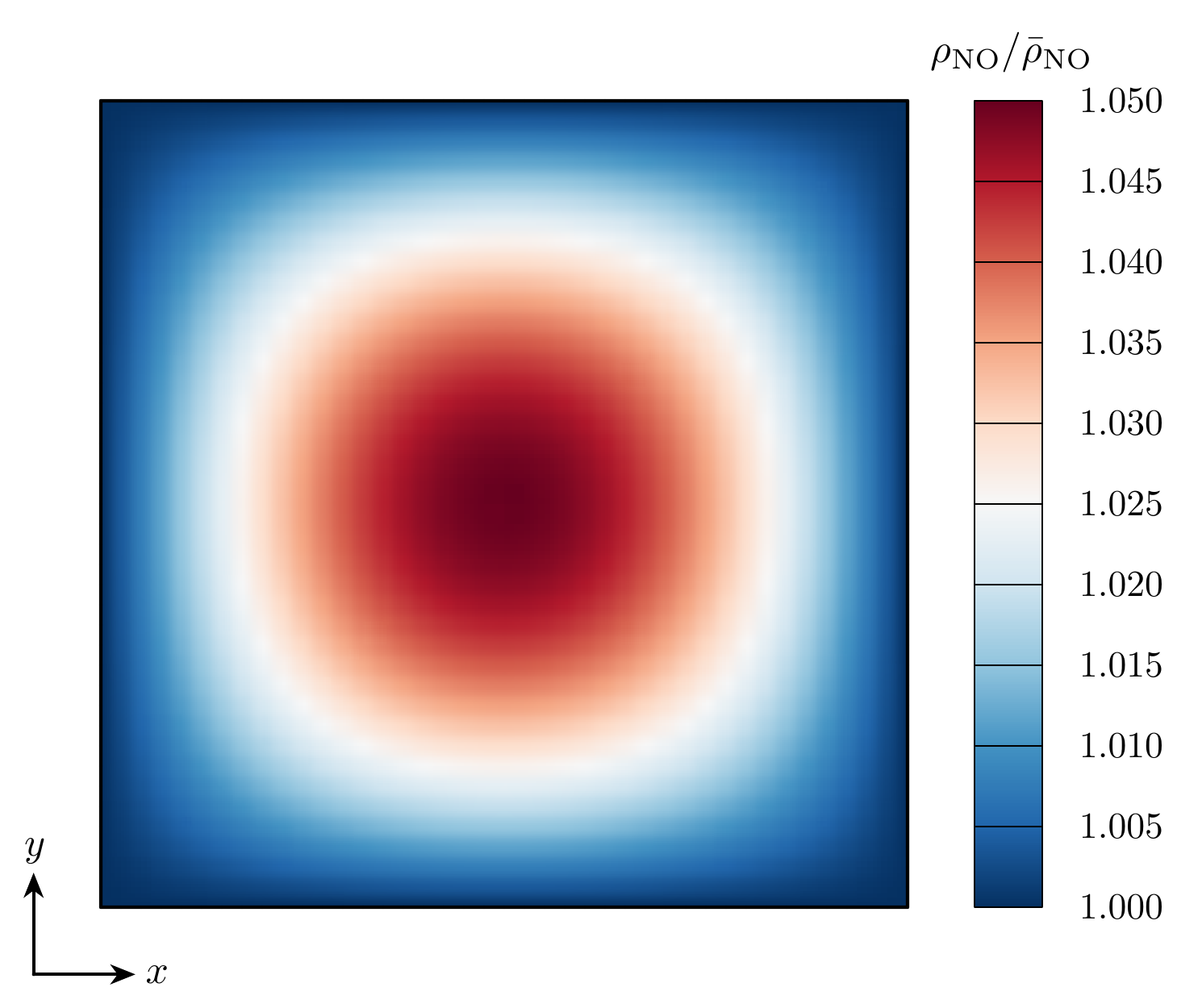} \\
\includegraphics[scale=.22,clip=true,trim=0in 0in 0in 0in]{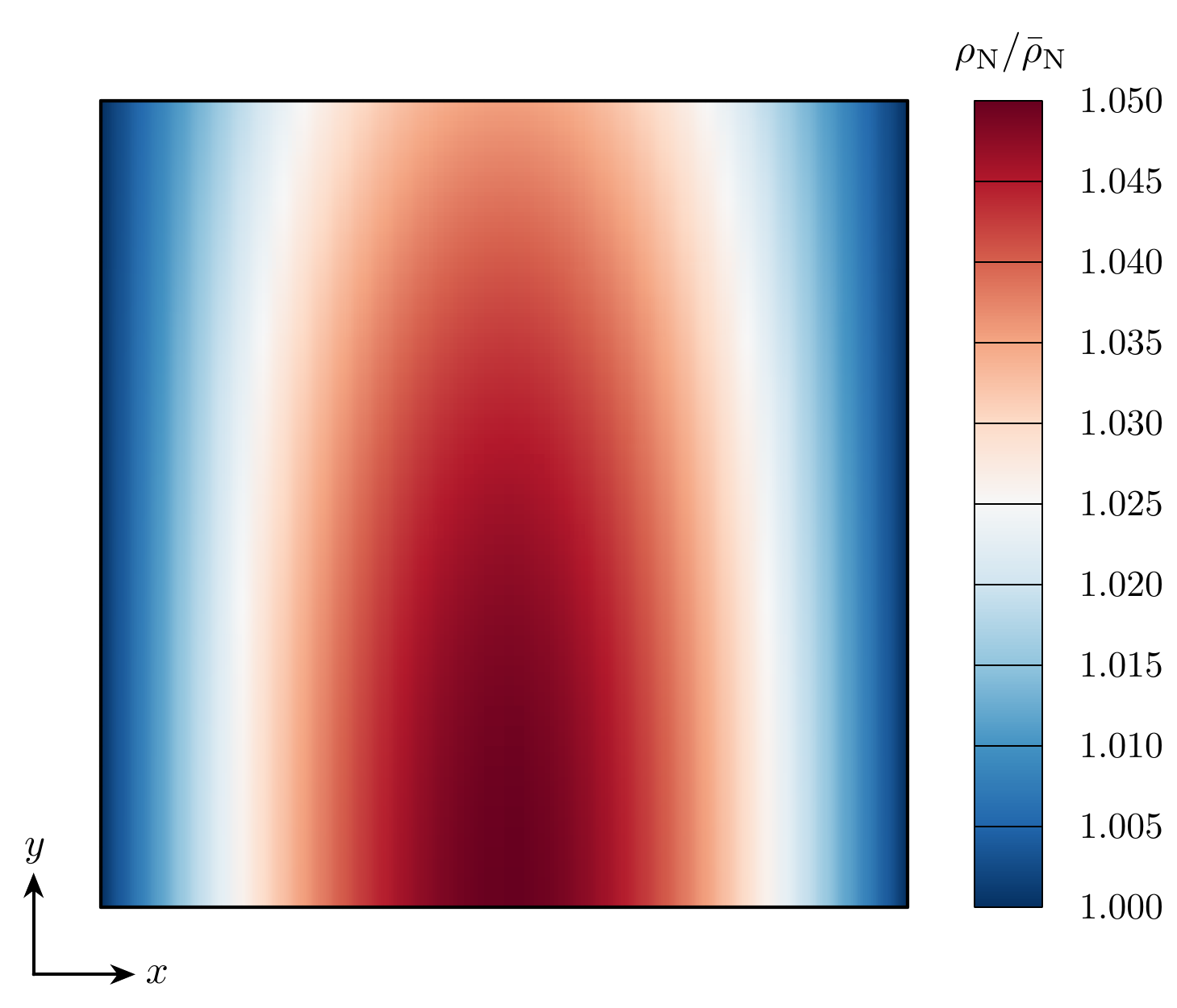}
\includegraphics[scale=.22,clip=true,trim=0in 0in 0in 0in]{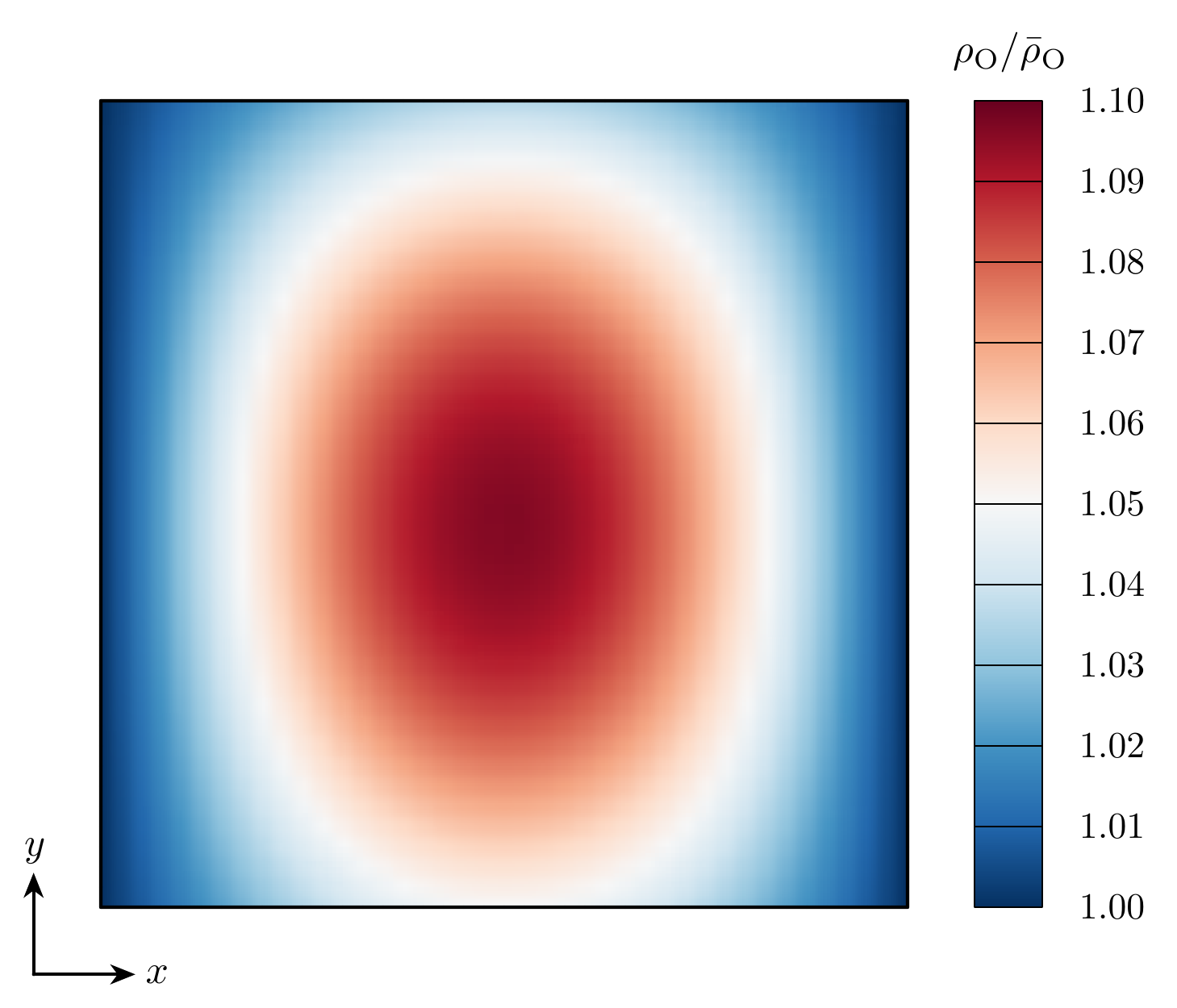}
\includegraphics[scale=.22,clip=true,trim=0in 0in 0in 0in]{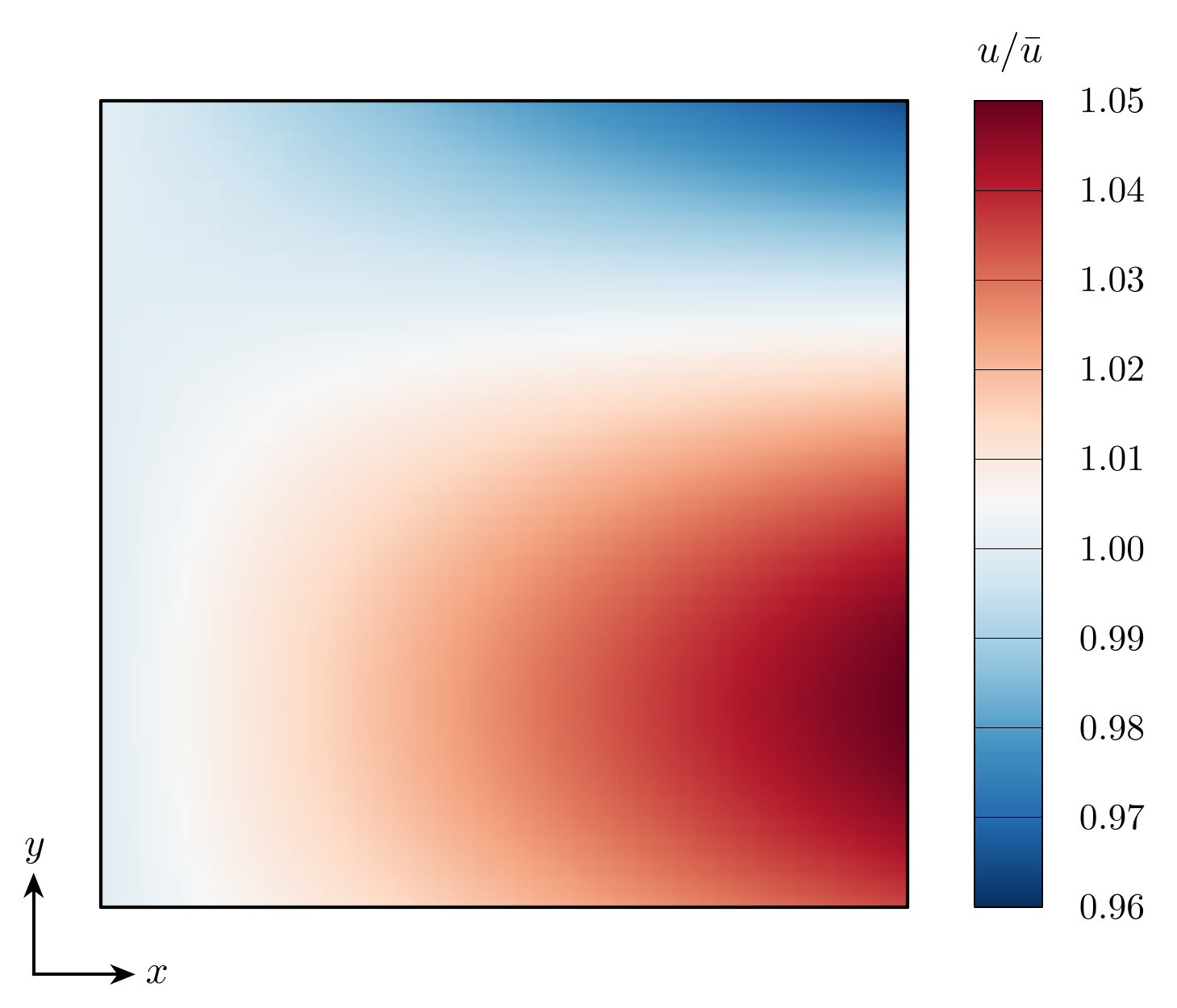} \\
\includegraphics[scale=.22,clip=true,trim=0in 0in 0in 0in]{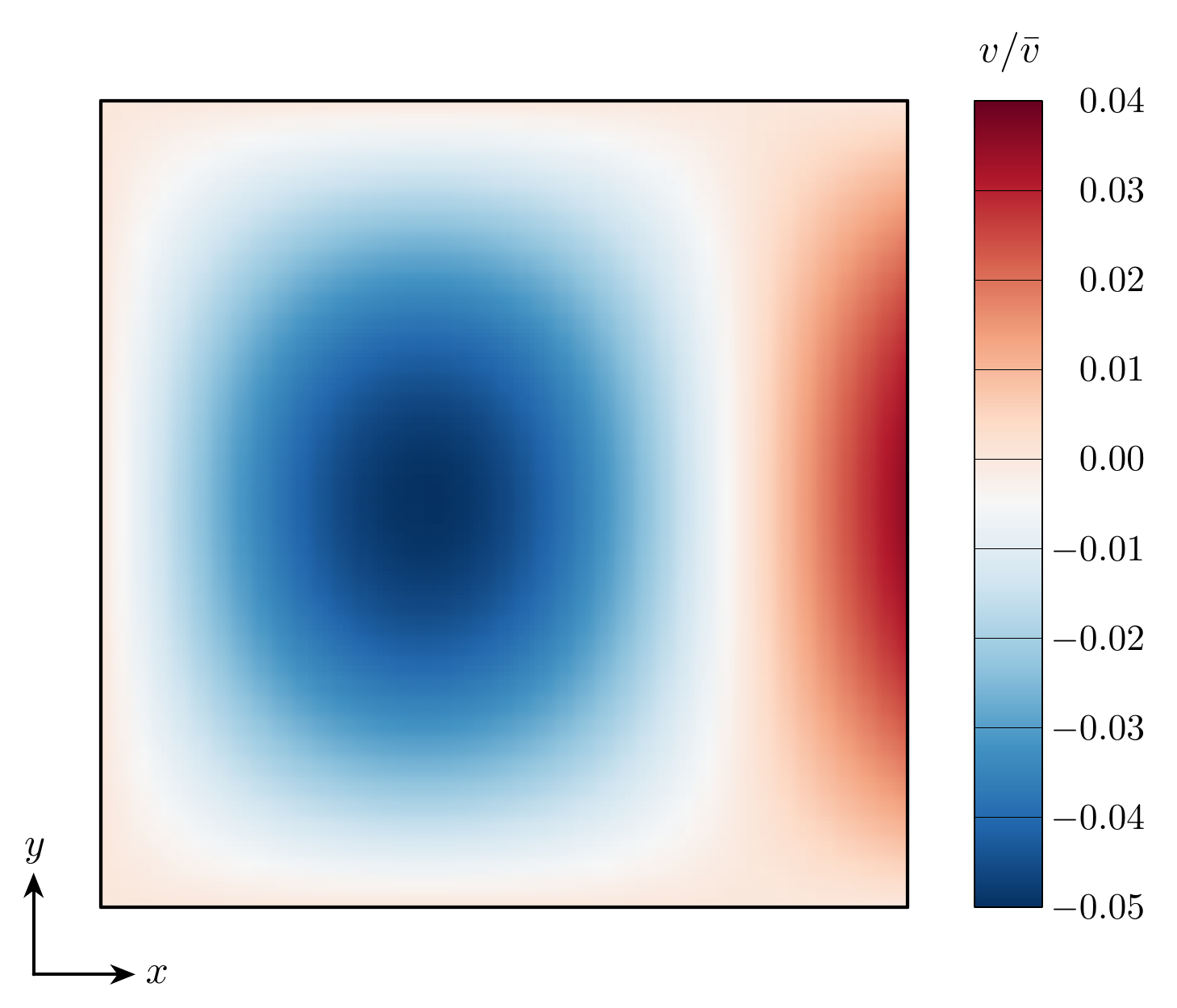}
\includegraphics[scale=.22,clip=true,trim=0in 0in 0in 0in]{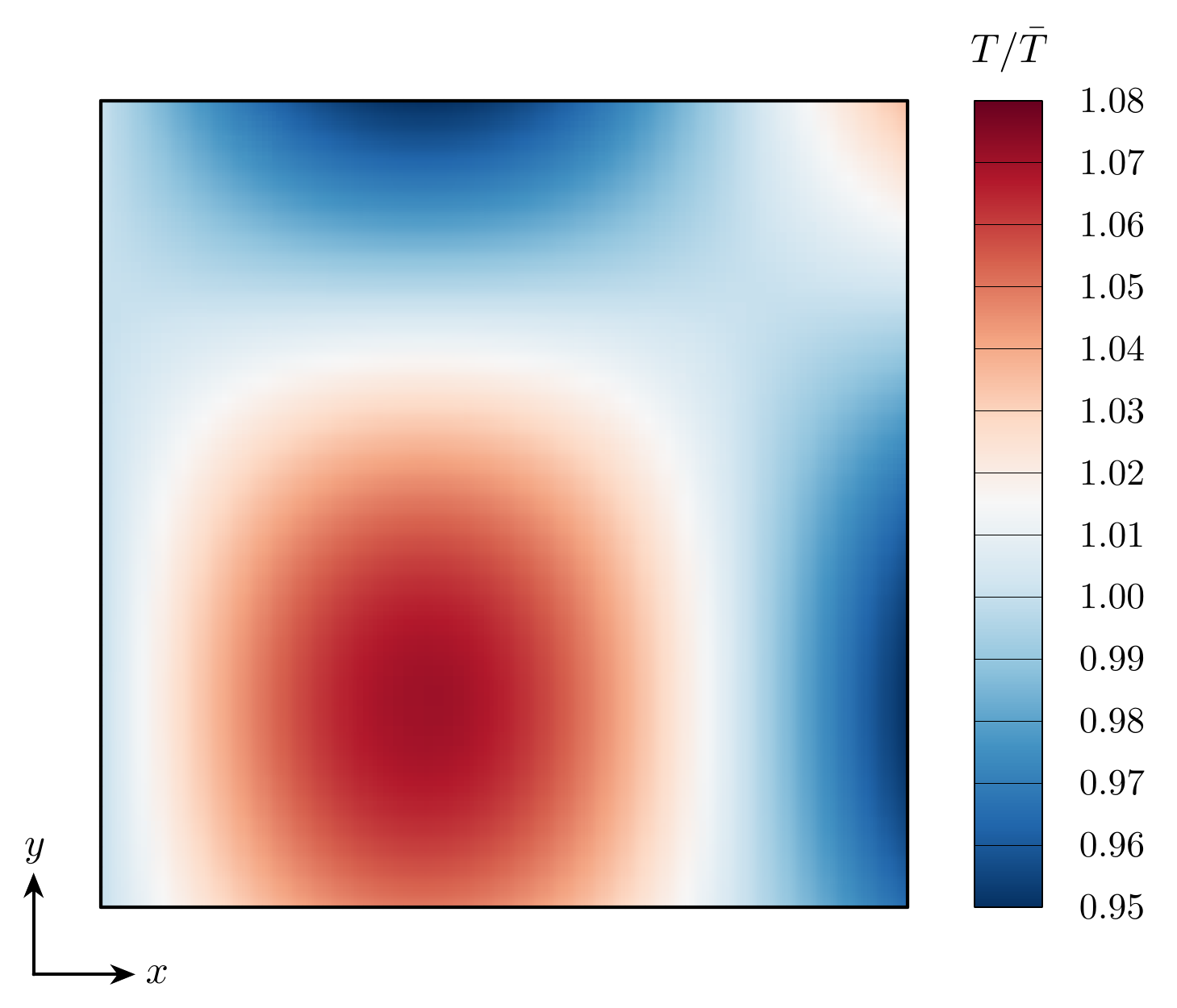}
\includegraphics[scale=.22,clip=true,trim=0in 0in 0in 0in]{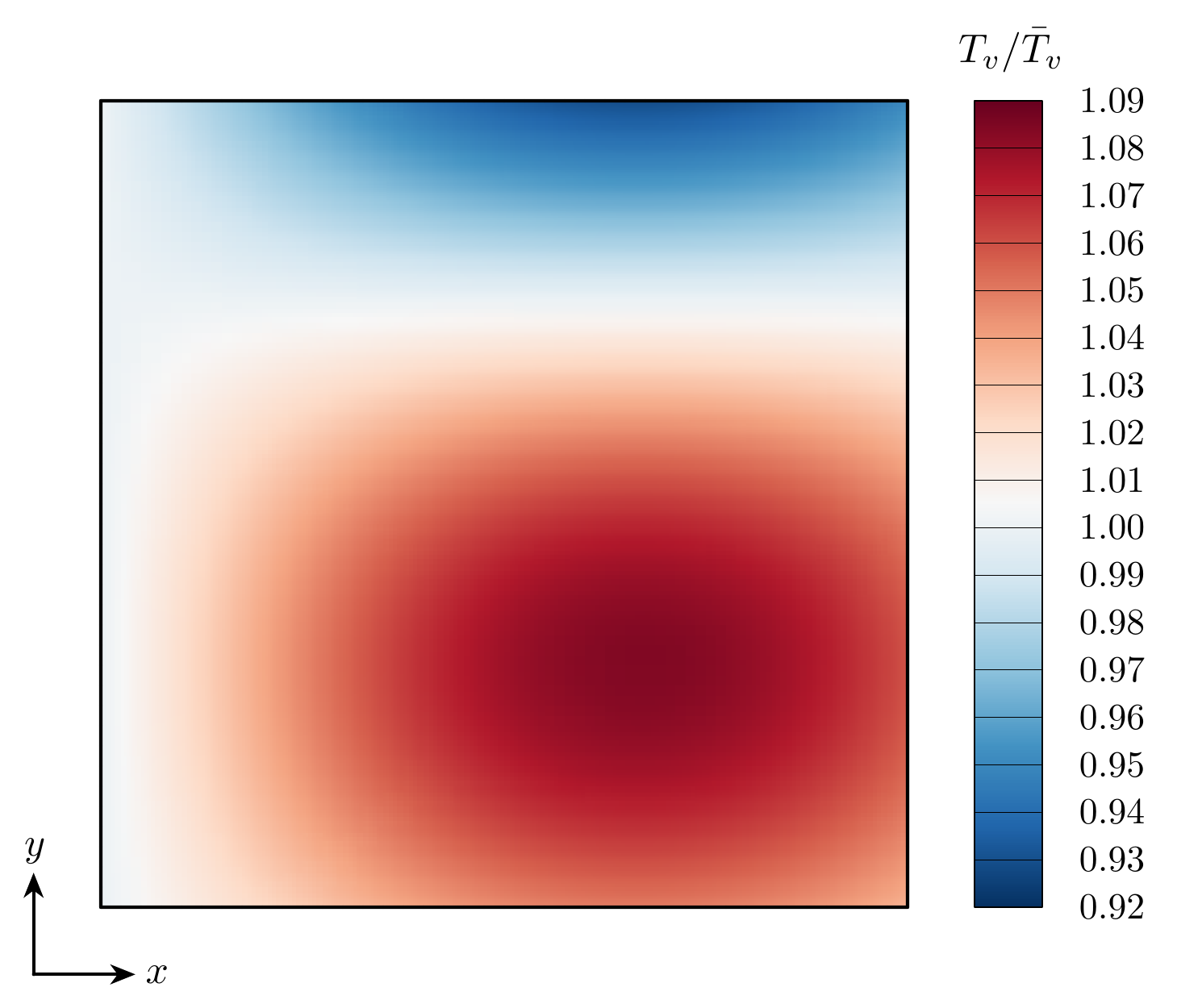}
\caption{2D manufactured solutions from Equation~\eqref{eq:2d_solutions} with $\perturbation=0.05$.}
\label{fig:2d_solutions}
\end{figure}

For many of our two-dimensional manufactured solutions, we use the following solution structures, or subsets thereof:
%
\begin{alignat}{8}
&\rho_{\text{N}_2}&&(x,y) = \bar{\rho}_{\text{N}_2}&&\bigl[1          - \perturbation\sin\left(         \tfrac{5}{4} \pi x\right)\bigl(\sin\left(\phantom{\tfrac{1}{1}}\pi y\right)+{}&\cos\left(\phantom{\tfrac{1}{1}}\pi y\right)\bigr)\bigr],
\nonumber\\[.5em]
&\rho_{\text{O}_2}&&(x,y) = \bar{\rho}_{\text{O}_2}&&\bigl[1          + \perturbation\sin\left(         \tfrac{3}{4} \pi x\right)\bigl(\sin\left(\phantom{\tfrac{1}{1}}\pi y\right)+{}&\cos\left(\phantom{\tfrac{1}{1}}\pi y\right)\bigr)\bigr],
\nonumber\\[.5em]
&\rho_{\text{NO}} &&(x,y) = \bar{\rho}_{\text{NO}} &&\bigl[1          + \perturbation\sin\left(\phantom{\tfrac{1}{1}}\pi x\right)\bigl(\sin\left(\phantom{\tfrac{1}{1}}\pi y\right)   &                                            \bigr)\bigr],
\nonumber\\[.5em]
&\rho_{\text{N}}  &&(x,y) = \bar{\rho}_{\text{N}}  &&\bigl[1          + \perturbation\sin\left(\phantom{\tfrac{1}{1}}\pi x\right)\bigl(\cos\left(         \tfrac{1}{4} \pi y\right)   &                                            \bigr)\bigr],
\nonumber\\[.5em]
&\rho_{\text{O}}  &&(x,y) = \bar{\rho}_{\text{O}}  &&\bigl[1          + \perturbation\sin\left(\phantom{\tfrac{1}{1}}\pi x\right)\bigl(\sin\left(\phantom{\tfrac{1}{1}}\pi y\right)+{}&\cos\left(         \tfrac{1}{4} \pi y\right)\bigr)\bigr],
\nonumber\\[.5em]
&u                &&(x,y) = \bar{u}                &&\bigl[1          + \perturbation\sin\left(         \tfrac{1}{4} \pi x\right)\bigl(\sin\left(\phantom{\tfrac{1}{1}}\pi y\right)+{}&\cos\left(\phantom{\tfrac{1}{1}}\pi y\right)\bigr)\bigr],
\nonumber\\[.5em]
&v                &&(x,y) = \bar{v}                &&\bigl[\phantom{0}- \perturbation\sin\left(         \tfrac{5}{4} \pi x\right)\bigl(\sin\left(\phantom{\tfrac{1}{1}}\pi y\right)   &                                            \bigr)\bigr],
\nonumber\\[.5em]
&T                &&(x,y) = \bar{T}                &&\bigl[1          + \perturbation\sin\left(         \tfrac{5}{4} \pi x\right)\bigl(\sin\left(\phantom{\tfrac{1}{1}}\pi y\right)+{}&\cos\left(\phantom{\tfrac{1}{1}}\pi y\right)\bigr)\bigr],
\nonumber\\[.5em]
&T_v              &&(x,y) = \bar{T}_v              &&\bigl[1          + \perturbation\sin\left(         \tfrac{3}{4} \pi x\right)\bigl(\sin\left(         \tfrac{5}{4} \pi y\right)+{}&\cos\left(         \tfrac{3}{4} \pi y\right)\bigr)\bigr].
\label{eq:2d_solutions}
\end{alignat}
These solutions are shown in Figure~\ref{fig:2d_solutions} for $(x,y)\in[0,\,1]\text{ m}\times [0,\,1]\text{ m}$.  The inflow and outflow boundaries are located at $x=0$ m and $x=1$ m.  The velocity field ensures the flow is tangential to the slip-wall (tangent-flow) boundaries located at $y=0$ m and $y=1$ m.

\subsection{Single-Species Inviscid Flow in Thermochemical Equilibrium} 

The first set of tests consists of a single species ($\nspecies=1$, $\rhovec=\rho$) inviscid ($\fluxd=0$) flow in thermochemical equilibrium ($\source=\mathbf{0}$, $T=T_v$).  For this type of flow, the reference velocity $\bar{\xvel}$ is determined from a reference Mach number $\bar{\mach}$:
\begin{align}
\bar{\xvel}=\bar{\mach}\sqrt{\gamma \rgas \bar{T}},
\label{eq:ubar}
\end{align}
where $\gamma$ is the ratio of specific heats for air, and $\rgas$ is the specific gas constant of air.

\subsubsection{1D Supersonic Flow using a Manufactured Solution} 
\def \prefix {1D MMS, $\nspecies=1$, $T_v=T$, $\wvecdot=\mathbf{0}$}
\def \preftex {onedinviscidmms}
\label{sec:\preftex}

For this test, we simulate a simple, one-dimensional flow using manufactured solutions:
\begin{alignat*}{4}
\rho(x) &{}={}& \bar{\rho}\left[1-\perturbation\sin(\pi x)\right],
\\
\xvel(x) &{}={}& \bar{\xvel}\left[1-\perturbation\sin(\pi x)\right],
\\
T(x) &{}={}& \bar{T}\left[1+\perturbation\sin(\pi x)\right],
\end{alignat*}
with $x\in[0,\,1]$ m, 
$\bar{\rho}=1$ kg/m$^3$,
$\bar{\mach}=2.5$,  
$\bar{T}=300$ K, 
and 
$\perturbation=0.05$.

Upon solving~\eqref{eq:mms}, the observed order of accuracy $\globalaccuracy$ is computed from \eqref{eq:accuracy_norm} for $\scalar=\{\rho,\,\xvel,\,T\}$.
Five 1D meshes are used, consisting of 50, 100, 200, 400, and 800 elements.

Table~\ref{tab:\preftex/orig} shows the observed order of accuracy from the original state of the code using the \linfnorm{} and \lonenorm{} of the error.  The \lonenorm{} indicates second-order accuracy $(\globalaccuracy=2)$, whereas the \linfnorm{} indicates first-order accuracy $(\globalaccuracy=1)$.  This example demonstrates the usefulness of the \linfnorm{}.  The \lonenorm{} suggests the code, on average, is second-order accurate; however, the \linfnorm{} captures localized deviations in the order of accuracy.  These deviations are due to the supersonic-inflow and supersonic-outflow boundary-condition implementations being only first-order accurate. For this case, the order reduction is limited to the vicinity of the boundaries.

\begin{table}[!b]
\centering
\begin{tabular}{c c c c c c c}
\toprule
& \multicolumn{3}{c}{\linfnorm{}} & \multicolumn{3}{c}{\lonenorm{}} \\
\cmidrule(lr){2-4} \cmidrule(lr){5-7}
Mesh & $\rho$ & $u$ & $T$ & $\rho$ & $u$ & $T$ \\
\midrule
1--2 & 1.0008 & 1.0008 & 1.0008 & 1.9955 & 1.9956 & 1.9956 \\
2--3 & 1.0002 & 1.0002 & 1.0002 & 1.9980 & 1.9981 & 1.9981 \\
3--4 & 1.0001 & 1.0001 & 1.0000 & 1.9994 & 1.9994 & 1.9995 \\
4--5 & 1.0000 & 1.0000 & 1.0000 & 1.9998 & 1.9998 & 1.9998 \\
\bottomrule
\end{tabular}
\caption{\prefix: Observed order of accuracy $\globalaccuracy$ from original boundary conditions.}
\label{tab:\preftex/orig}
\end{table}

We corrected the two boundary-condition implementations to be second-order accurate, which is confirmed in Table~\ref{tab:\preftex/corr}.  Additionally, Figures~\ref{fig:\preftex/linf} and~\ref{fig:\preftex/lone} show the two error norms for each of the flow variables, before and after correcting the boundary conditions.  As shown in Figure~\ref{fig:\preftex/linf}, the maximum errors are reduced by orders of magnitude upon correcting the boundary conditions.  Furthermore, Figure~\ref{fig:\preftex/lone} shows that the correct boundary conditions reduce the average error by a factor of approximately three.

\begin{table}
\centering
\begin{tabular}{c c c c c c c}
\toprule
& \multicolumn{3}{c}{\linfnorm{}} & \multicolumn{3}{c}{\lonenorm{}} \\
\cmidrule(lr){2-4} \cmidrule(lr){5-7}
Mesh & $\rho$ & $u$ & $T$ & $\rho$ & $u$ & $T$ \\
\midrule
1--2 & 2.0313 & 2.0362 & 2.0351 & 2.0489 & 2.0526 & 2.0521 \\
2--3 & 2.0157 & 2.0184 & 2.0178 & 2.0252 & 2.0271 & 2.0268 \\
3--4 & 2.0079 & 2.0093 & 2.0090 & 2.0128 & 2.0138 & 2.0136 \\
4--5 & 2.0040 & 2.0047 & 2.0045 & 2.0064 & 2.0070 & 2.0069 \\
\bottomrule
\end{tabular}
\caption{\prefix: Observed order of accuracy $\globalaccuracy$ from corrected boundary conditions.}
\label{tab:\preftex/corr}
\end{table}

\begin{figure}
\centering%
\begin{subfigure}[t]{0.49\columnwidth}\includegraphics[scale=.64,clip=true,trim=2.2in 0.05in  2.8in 0.15in]{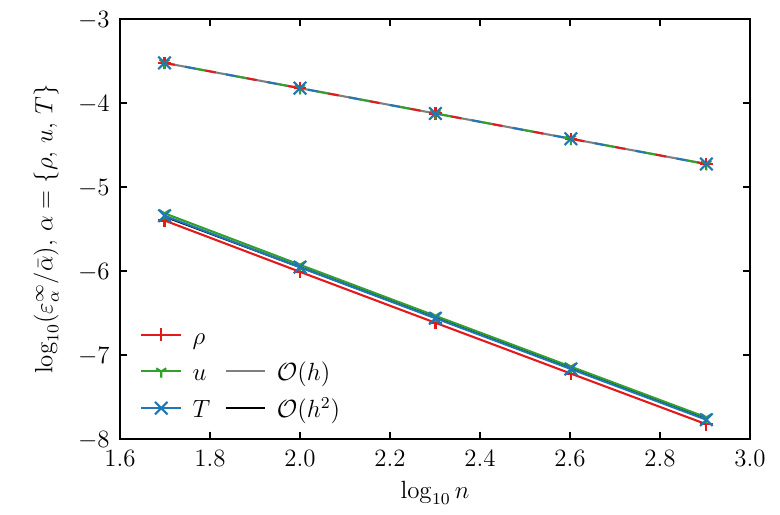}\caption{\linfnorm{}}\label{fig:\preftex/linf}\end{subfigure} \hfill
\begin{subfigure}[t]{0.49\columnwidth}\includegraphics[scale=.64,clip=true,trim=2.2in 0.05in  2.8in 0.15in]  {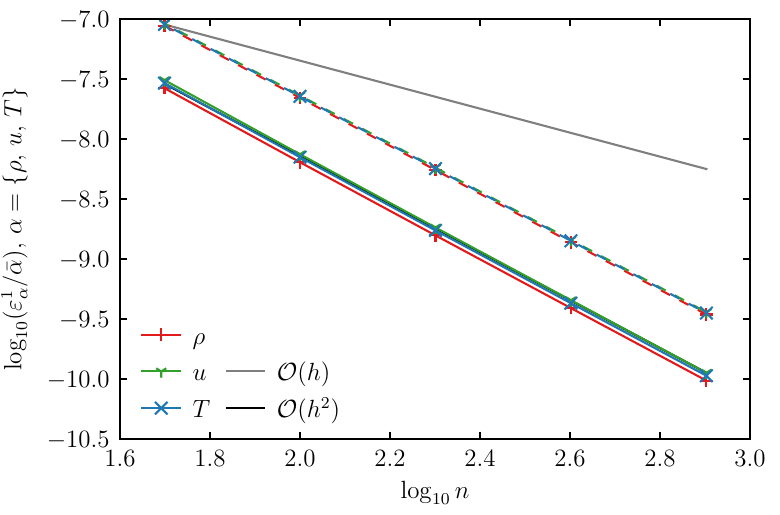}\caption{\lonenorm{}}\label{fig:\preftex/lone}\end{subfigure}
\caption{\prefix: Norms of the error.  Dashed lines denote original boundary conditions; solid lines denote corrected boundary conditions.}
\label{fig:\preftex/norms}
\end{figure}

\FloatBarrier   
\subsubsection{2D Supersonic Flow using a Manufactured Solution} 
\label{subsubsec:twodinviscidmms}

\def \prefix {2D MMS, $\nspecies=1$, $T_v=T$, $\wvecdot=\mathbf{0}$}
\def \preftex {twodinviscidmms}

This test increases the complexity of the first test by including variations along a second dimension.  In addition to the supersonic-inflow and supersonic-outflow boundary conditions of the first test, the slip-wall boundary condition is exercised.

The manufactured solutions for this case for $\scalar=\{\rho,\,\xvel,\,\yvel,\,T\}$ are listed in~\eqref{eq:2d_solutions} and shown in Figure~\ref{fig:2d_solutions}.  
In~\eqref{eq:2d_solutions} and in Figure~\ref{fig:2d_solutions}, $\rho=\rho_{\text{N}_2}$, $\bar{\rho}=\bar{\rho}_{\text{N}_2}$, and $\bar{\yvel}=\bar{\xvel}$~\eqref{eq:ubar}.
The domain is a square with $(x,y)\in[0,\,1]\text{ m}\times [0,\,1]\text{ m}$, and 
$\bar{\rho}=1$ kg/m$^3$,
$\bar{\mach}=2.5$,  
$\bar{T}=300$ K, 
and 
$\perturbation=0.05$.

Upon solving~\eqref{eq:mms}, the observed order of accuracy $\globalaccuracy$ is computed from \eqref{eq:accuracy_norm}.
Five 2D meshes are used, consisting of $25\times 25$, $50\times 50$, $100\times 100$, $200\times 200$, and $400\times 400$ elements.  These meshes are chosen to test the spatial accuracy of the discretization for nonuniform meshes, \reviewerOne{and are created using the approach in~\ref{appx_b}}.  The $50\times 50$ mesh is shown in Figure~\ref{fig:\preftex/mesh}.

Table~\ref{tab:\preftex/orig} shows the observed order of accuracy using the \linfnorm{} and \lonenorm{} of the error.  Both norms indicate first-order accuracy $(\globalaccuracy=1)$, despite the second-order-accuracy expectation.  This inconsistency is due to the supersonic-inflow, supersonic-outflow, and slip-wall boundary-condition implementations being only first-order accurate.  Unlike the first case, the implications of the first-order-accurate boundary conditions are global for this case.

The corrected boundary-condition implementations are confirmed to be second-order accurate $(\globalaccuracy=2)$ in Table~\ref{tab:\preftex/corr}.  Figures~\ref{fig:\preftex/linf} and~\ref{fig:\preftex/lone} show the two error norms for each of the flow variables, before and after correcting the boundary conditions.  The correct boundary conditions reduce both the maximum and average error by orders of magnitude.  For the subsequent results, we omit the \lonenorm{} and consider only the corrected boundary conditions.

\begin{figure}
\centering
\includegraphics[scale=.33,clip=true,trim=0in 0in 1.5in 0.5in]{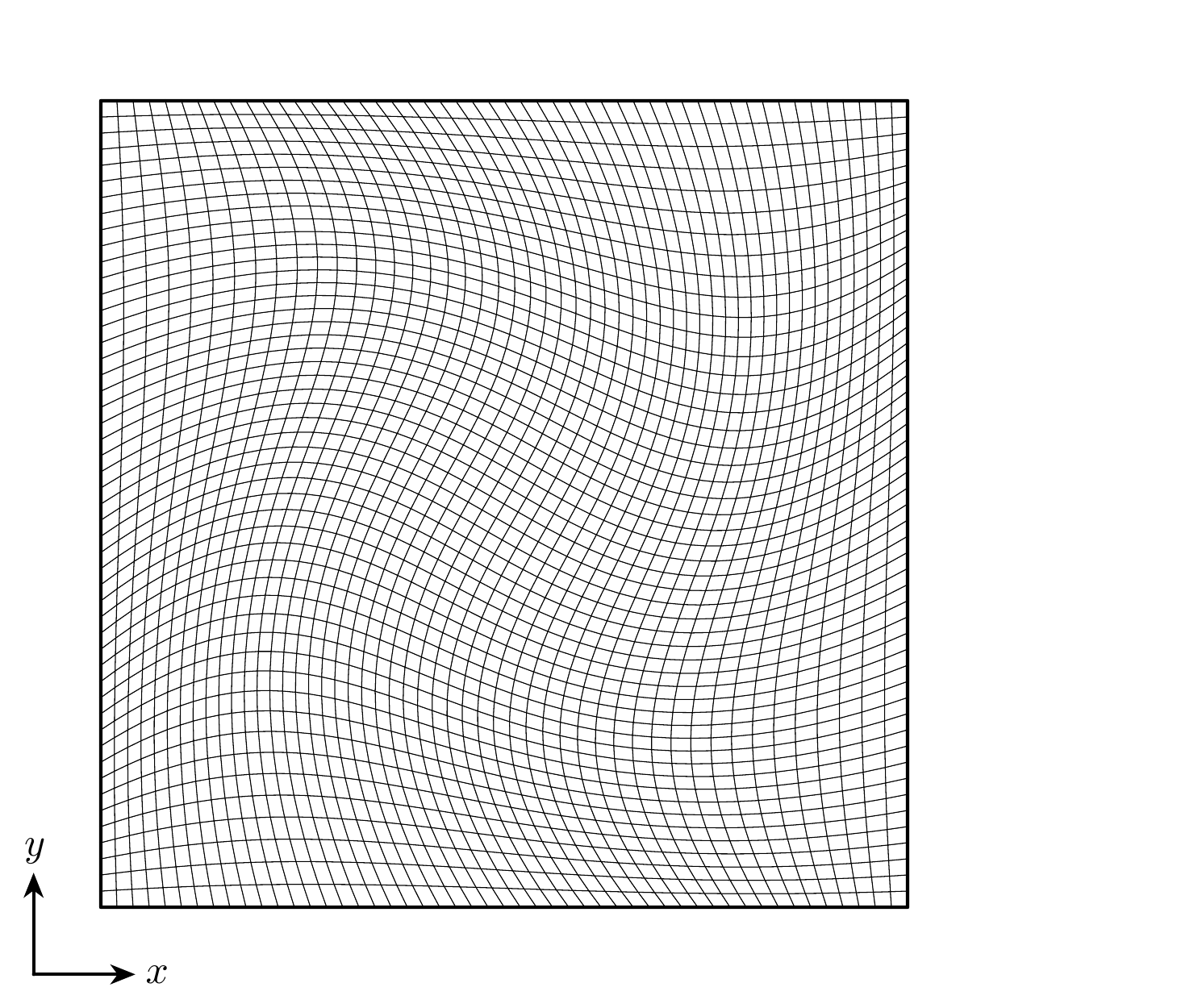}
\caption{\prefix: Mesh with $50\times 50$ elements.}
\label{fig:\preftex/mesh}
\end{figure}

\begin{table}[!t]
\centering
\begin{tabular}{c c c c c c c c c}
\toprule
& \multicolumn{4}{c}{\linfnorm{}} & \multicolumn{4}{c}{\lonenorm{}} \\
\cmidrule(lr){2-5} \cmidrule(lr){6-9}
Mesh & $\rho$ & $u$ & $v$ & $T$ & $\rho$ & $u$ & $v$ & $T$ \\
\midrule
1--2 & 0.9420 & 0.9409 & 0.9721 & 0.9628 & 1.0083 & 0.9838 & 0.9255 & 0.9861 \\
2--3 & 0.9850 & 0.9902 & 0.9910 & 0.9874 & 1.0082 & 0.9980 & 0.9686 & 0.9957 \\
3--4 & 0.9960 & 1.0002 & 0.9924 & 0.9952 & 1.0043 & 1.0008 & 0.9871 & 0.9985 \\
4--5 & 0.9989 & 1.0009 & 0.9959 & 0.9984 & 1.0022 & 1.0008 & 0.9943 & 0.9995 \\
\bottomrule
\end{tabular}
\caption{\prefix: Observed order of accuracy $\globalaccuracy$ from original boundary conditions.}
\label{tab:\preftex/orig}
\end{table}

\begin{table}[!t]
\centering
\begin{tabular}{c c c c c c c c c}
\toprule
& \multicolumn{4}{c}{\linfnorm{}} & \multicolumn{4}{c}{\lonenorm{}} \\
\cmidrule(lr){2-5} \cmidrule(lr){6-9}
Mesh & $\rho$ & $u$ & $v$ & $T$ & $\rho$ & $u$ & $v$ & $T$ \\
\midrule
1--2 & 2.0623 & 1.9188 & 1.8174 & 1.8598 & 2.2440 & 2.1789 & 2.1000 & 2.1802 \\
2--3 & 2.1304 & 1.9450 & 1.9221 & 1.9280 & 2.1701 & 2.1248 & 2.0745 & 2.1038 \\
3--4 & 2.0902 & 1.9603 & 1.9671 & 1.9586 & 2.0788 & 2.0577 & 2.0436 & 2.0461 \\
4--5 & 2.0128 & 1.9823 & 1.9860 & 1.9809 & 2.0303 & 2.0246 & 2.0230 & 2.0220 \\
\bottomrule
\end{tabular}
\caption{\prefix: Observed order of accuracy $\globalaccuracy$ from corrected boundary conditions.}
\label{tab:\preftex/corr}
\end{table}

\begin{figure}
\centering%
\begin{subfigure}[t]{0.49\columnwidth}\includegraphics[scale=.64,clip=true,trim=2.2in 0.05in  2.8in 0.15in]{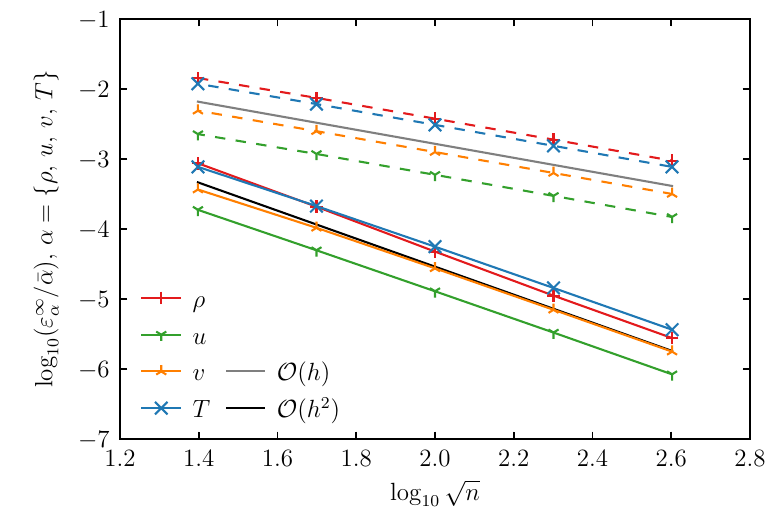}\caption{\linfnorm{}}\label{fig:\preftex/linf}\end{subfigure}\hfill
\begin{subfigure}[t]{0.49\columnwidth}\includegraphics[scale=.64,clip=true,trim=2.2in 0.05in  2.8in 0.15in]  {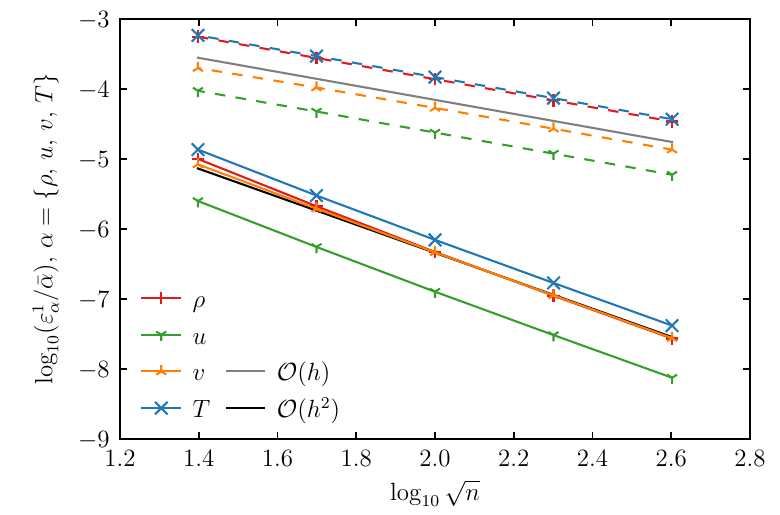}\caption{\lonenorm{}}\label{fig:\preftex/lone}\end{subfigure}
\caption{\prefix: Norms of the error.  Dashed lines denote original boundary conditions; solid lines denote corrected boundary conditions.}
\label{fig:\preftex/norms}
\end{figure}

\FloatBarrier

\subsubsection{2D Supersonic Flow using an Exact Solution} 
\def \prefix {2D Exact, $\nspecies=1$, $T_v=T$, $\wvecdot=\mathbf{0}$}
\def \preftex {twodinviscidexact}

This test exercises the same boundary conditions exercised in Section~\ref{subsubsec:twodinviscidmms}, but for an exact solution that does not require an additional source term.  The exact solution is a steady, isentropic vortex, which we simulate in a quarter-annulus domain~\cite{aftosmis_1995,luo_1995,krivodonova_2006}.

\begin{figure}[!th]
\vspace{-.5em}
\centering
\setlength{\fboxsep}{0pt}
\includegraphics[scale=.275,clip=true,trim=.4in .5in .25in .5in]{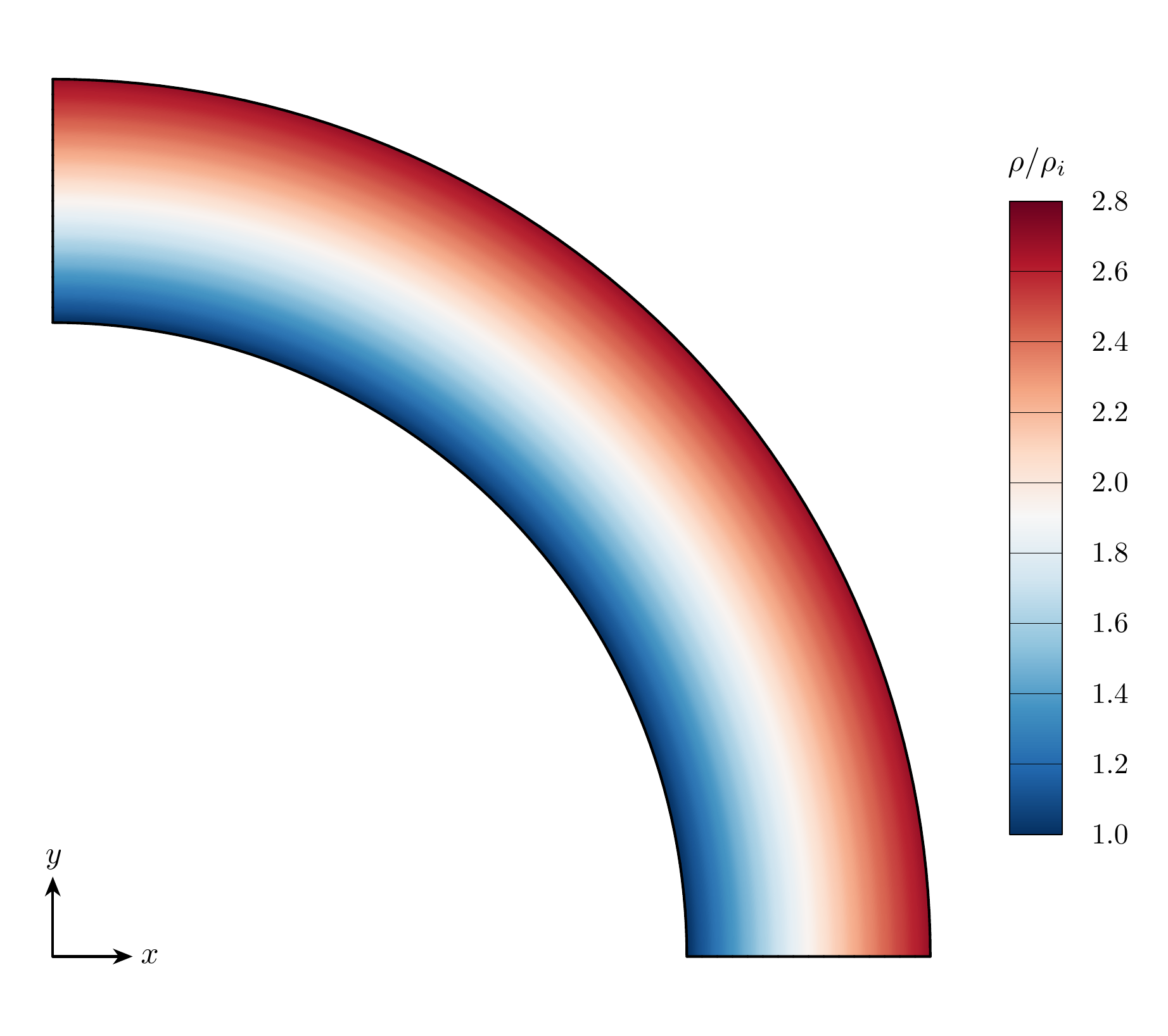}
\includegraphics[scale=.275,clip=true,trim=.4in .5in .25in .5in]{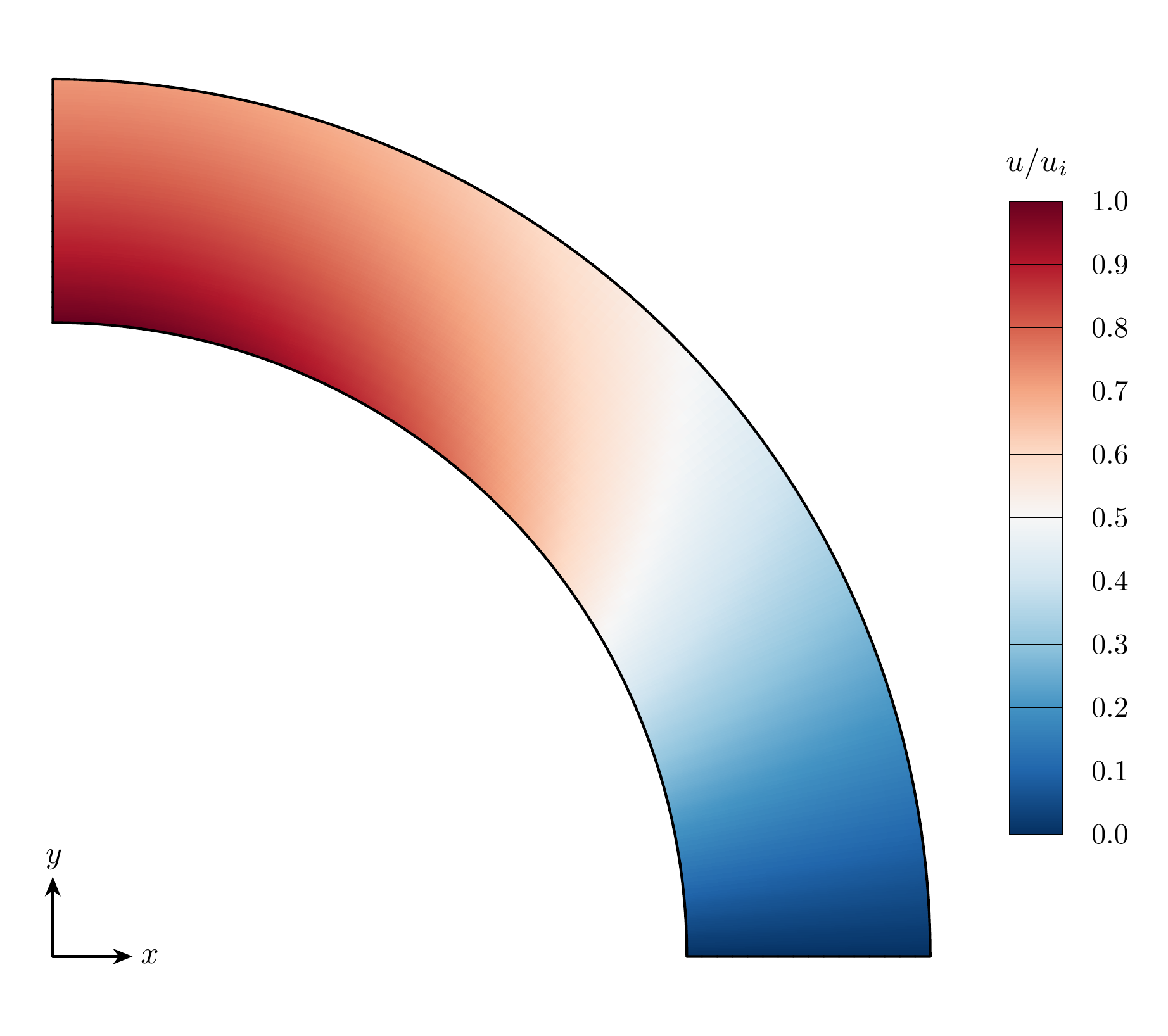} \\
\includegraphics[scale=.275,clip=true,trim=.4in .5in .25in .5in]{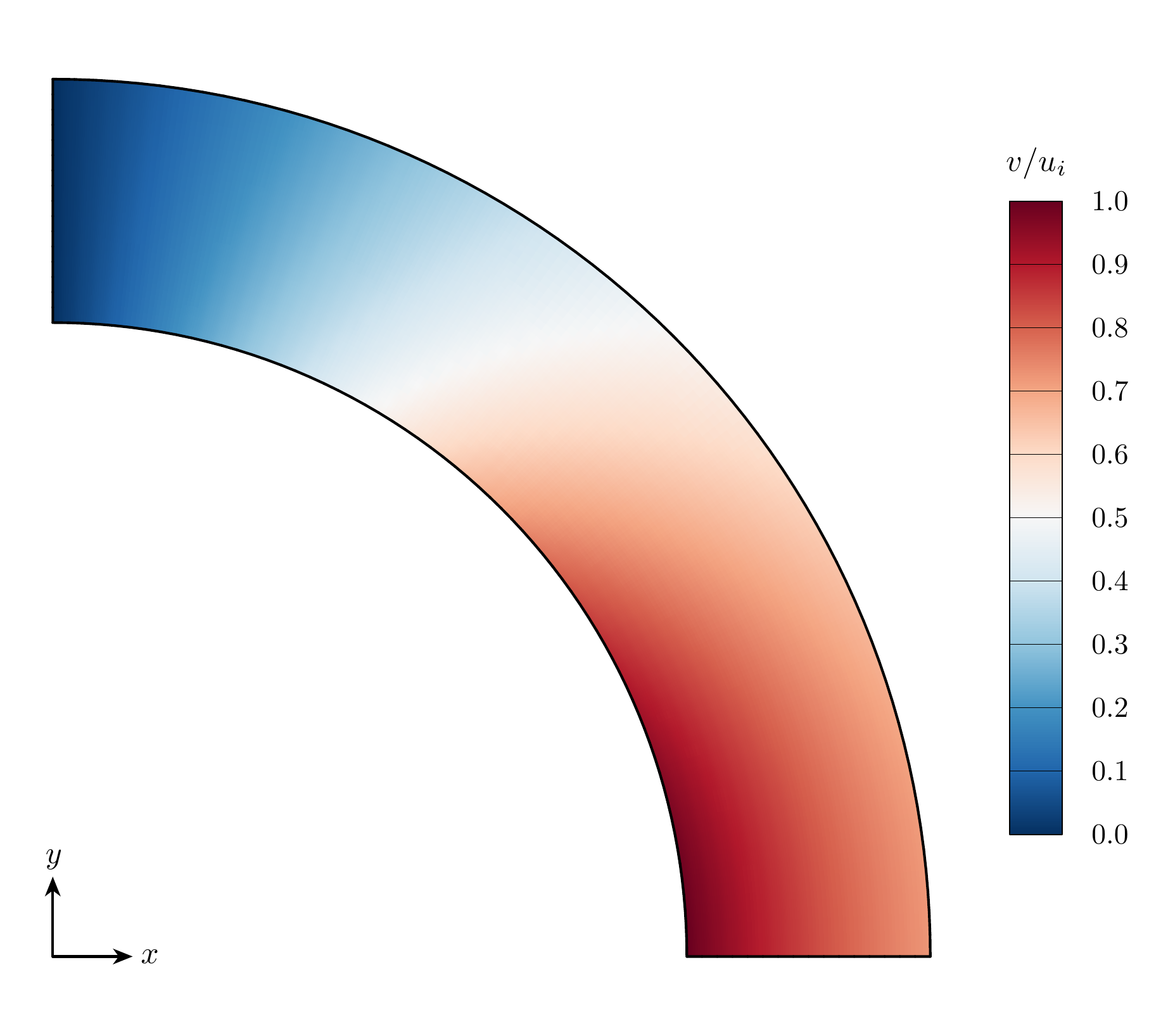}
\includegraphics[scale=.275,clip=true,trim=.4in .5in .25in .5in]{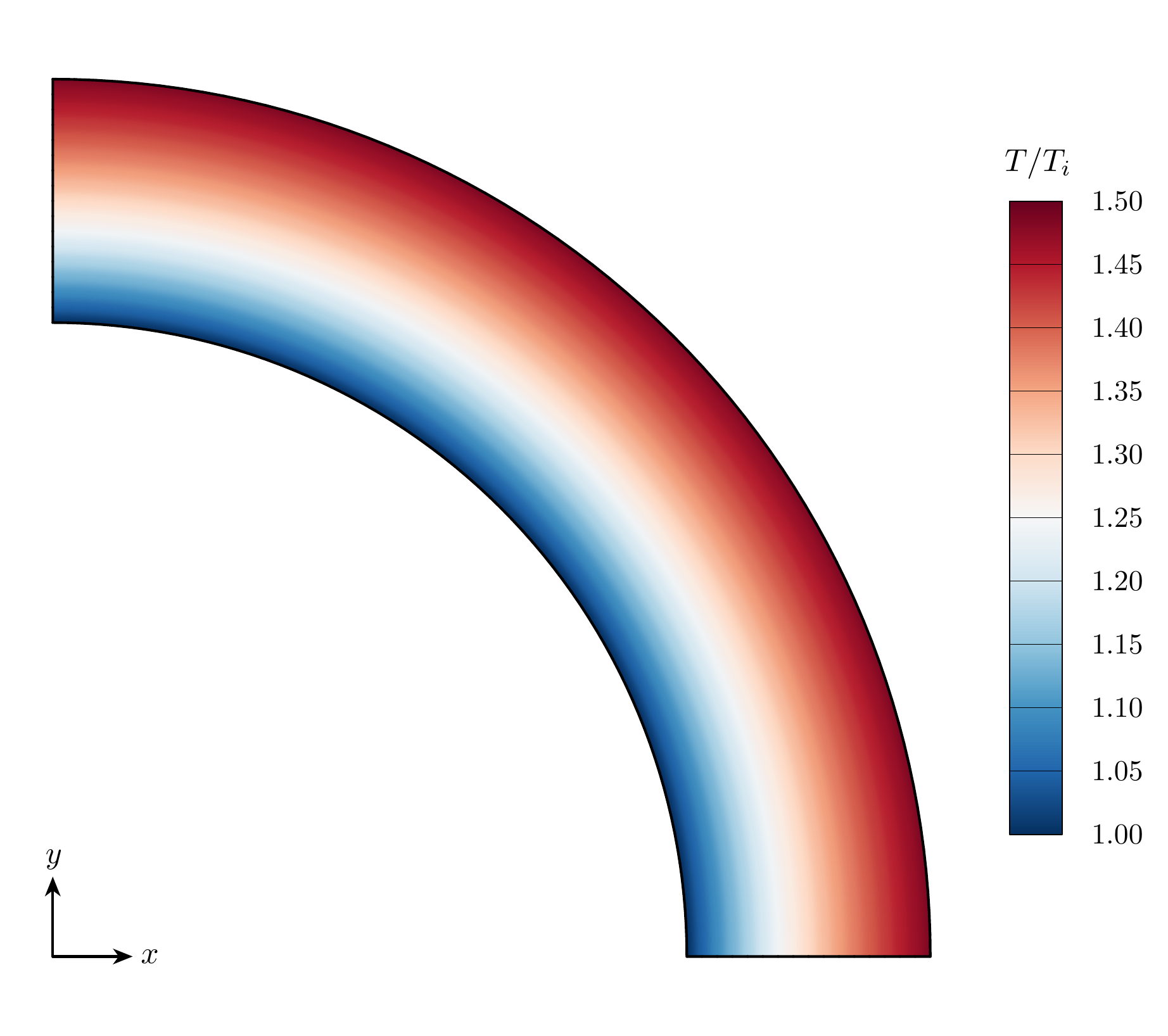}
\caption{\prefix: Exact solutions.}
\label{fig:\preftex/solution}
\end{figure}

The exact solutions for this case are
\begin{align*}
\rho(r) &{}= \rho_i\left[1+\frac{\gamma-1}{2}M_i^2\left(1-\left(\frac{r_i}{r}\right)^2\right)\right]^{\frac{1}{\gamma-1}},
\\[.5em]
u_r(r) &{}= 0,
\\[.5em]
u_\theta(r) &{}= -\soundspeed_i M_i \frac{r_i}{r},
\\[.5em]
T(r) &{}= T_i\left[1+\frac{\gamma-1}{2}M_i^2\left(1-\left(\frac{r_i}{r}\right)^2\right)\right],
\end{align*}
with $\rho_i=1$, $\soundspeed_i=1$, $\mach_i=2.25$, and $T_i=1/(\gamma \rgas)$.  $r$ is the distance from the center of the full annulus, which is bounded between $r_i=1$ and $r_o=1.384$.  The solutions are shown in Figure~\ref{fig:\preftex/solution}, with $\xvel_i=\yvel_i=\soundspeed_i\mach_i$.

Upon solving~\eqref{eq:disc}, the observed order of accuracy $\globalaccuracy$ is computed from \eqref{eq:accuracy_norm} for $\scalar=\{\rho,\,\xvel,\,\yvel,\,T\}$.
Six 2D meshes are used, consisting of $32\times 8$, $64\times 16$, $128\times 32$, $256\times 64$, $512\times 128$, and $1024\times 256$ elements.  The $64\times 16$ mesh is shown in Figure~\ref{fig:\preftex/mesh}.

\begin{figure}[!h]
\centering
\includegraphics[scale=.275,clip=true,trim=0in 0in 1.5in 0.5in]{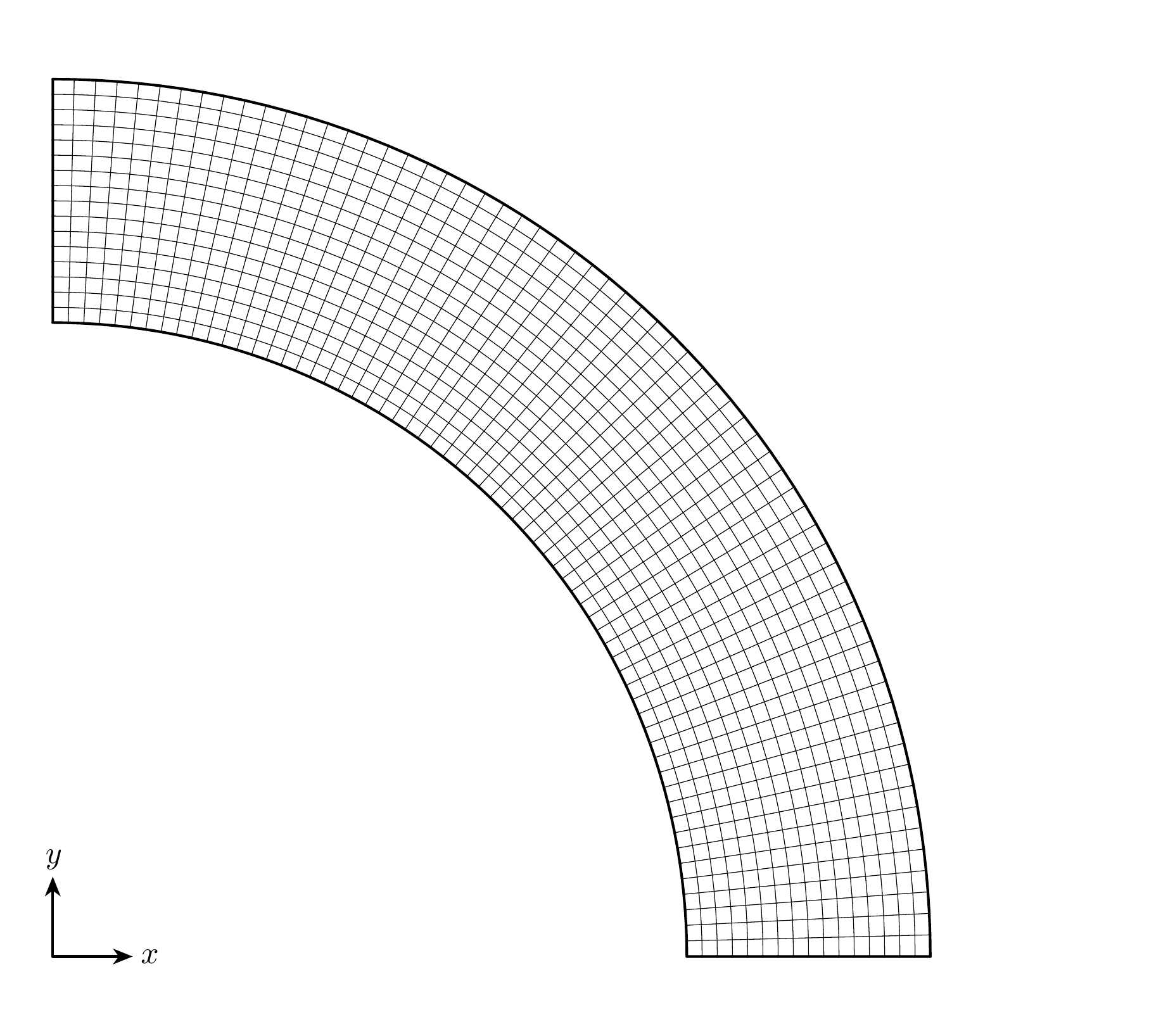}
\caption{\prefix: Mesh with $64\times 16$ elements.}
\label{fig:\preftex/mesh}
\end{figure}

Table~\ref{tab:\preftex/corr} shows the observed order of accuracy, using the \linfnorm{} of the error, which indicates second-order accuracy $(\globalaccuracy=2)$.  
Figure~\ref{fig:\preftex/linf} shows the \linfnorm{} for each of the flow variables.  
\begin{table}
\centering
\begin{tabular}{c c c c c}
\toprule
Mesh & $\rho$ & $u$ & $v$ & $T$\\
\midrule
1--2 & 1.9896 & 1.9119 & 1.9943 & 1.9699 \\
2--3 & 1.9735 & 1.9589 & 2.0070 & 1.9979 \\
3--4 & 1.9954 & 1.9760 & 2.0099 & 2.0076 \\
4--5 & 1.9972 & 1.9879 & 2.0054 & 2.0044 \\
5--6 & 1.9986 & 1.9940 & 2.0029 & 2.0025 \\
\bottomrule
\end{tabular}
\caption{\prefix: Observed order of accuracy $\globalaccuracy$ using \linfnorm{} of the error.}
\label{tab:\preftex/corr}
\end{table}

\FloatBarrier

\begin{figure}[!t]
\centering
\includegraphics[scale=.64,clip=true,trim=1.2in 0in 1.2in 0in]{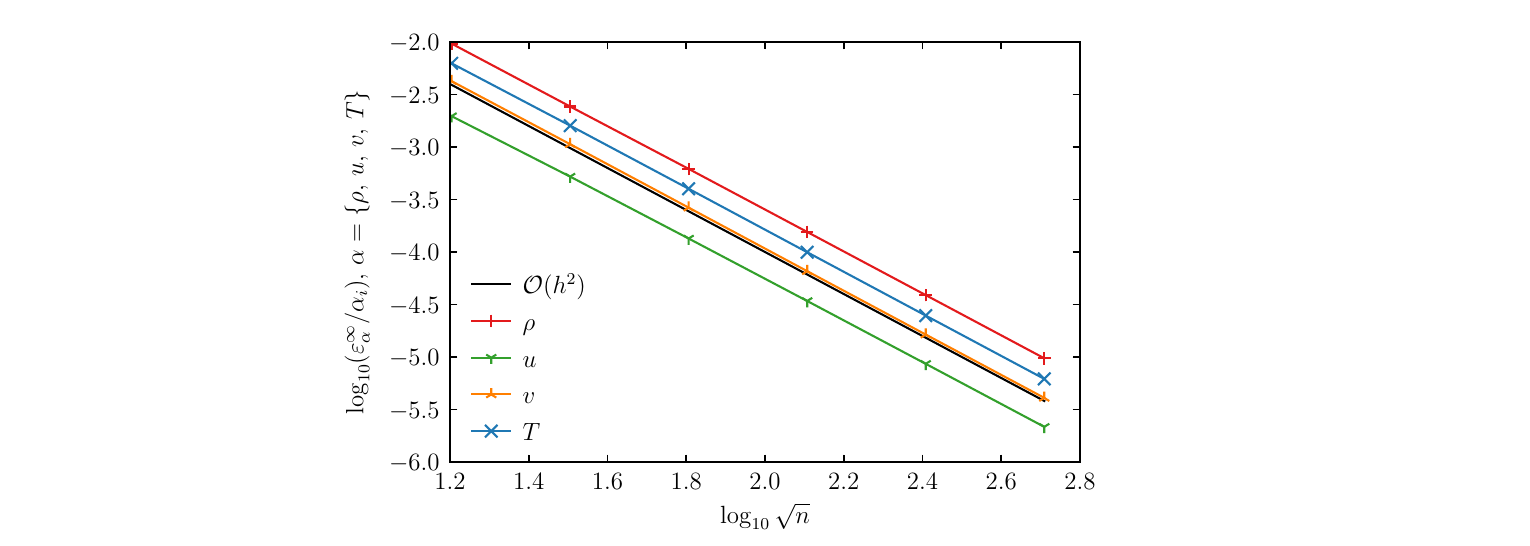}
\caption{\prefix: \linfnorm{} of the error.}
\label{fig:\preftex/linf}
\end{figure}

\subsubsection{3D Supersonic Flow using a Manufactured Solution} 
\def \prefix {3D MMS, $\nspecies=1$, $T_v=T$, $\wvecdot=\mathbf{0}$}
\def \preftex {threedinviscidmms}

For this test, we consider a three-dimensional flow.  The manufactured solutions for this case are
\begin{alignat*}{8}
 \rho(x,y,z) &&{}= \bar{\rho} &\bigl[1           - \perturbation\sin\left(\tfrac{5}{4}\pi x\right)(\sin(\pi y)+{}&\cos(\pi y))(&\sin(\pi z)+{}&\cos(\pi z))\bigr], \\[.5em]
\xvel(x,y,z) &&{}= \bar{\xvel}&\bigl[1           + \perturbation\sin\left(\tfrac{1}{4}\pi x\right)(\sin(\pi y)+{}&\cos(\pi y))(&\sin(\pi z)+{}&\cos(\pi z))\bigr], \\[.5em]
\yvel(x,y,z) &&{}= \bar{\yvel}&\bigl[\phantom{0} - \perturbation\sin\left(\tfrac{5}{4}\pi x\right)(\sin(\pi y) {}&           )(&\sin(\pi z)+{}&\cos(\pi z))\bigr], \\[.5em]
\zvel(x,y,z) &&{}= \bar{\zvel}&\bigl[\phantom{0} - \perturbation\sin\left(\tfrac{5}{4}\pi x\right)(\sin(\pi y)+{}&\cos(\pi y))(&\sin(\pi z) {}&           )\bigr], \\[.5em]
    T(x,y,z) &&{}= \bar{T}    &\bigl[1           + \perturbation\sin\left(\tfrac{5}{4}\pi x\right)(\sin(\pi y)+{}&\cos(\pi y))(&\sin(\pi z)+{}&\cos(\pi z))\bigr],
\end{alignat*}
with $(x,y,z)\in[0,\,1]\text{ m}\times [0,\,1]\text{ m}\times [0,\,1]\text{ m}$, and $\bar{\rho}=1$ kg/m$^3$,
$\bar{\mach}=2.5$,  
$\bar{\yvel}=\bar{\zvel}=\bar{\xvel}$~\eqref{eq:ubar},
$\bar{T}=300$ K, 
and 
$\perturbation=0.05$.  

Upon solving~\eqref{eq:mms}, the observed order of accuracy $\globalaccuracy$ is computed from \eqref{eq:accuracy_norm} for $\scalar=\{\rho,\,\xvel,\,\yvel,\,\zvel,\,T\}$.
Five 3D meshes are used, consisting of $25\times 25\times 25$, $50\times 50\times 50$, $100\times 100\times 100$, $200\times 200\times 200$, and $400\times 400\times 400$ elements.  These meshes are chosen to test the spatial accuracy of the discretization for nonuniform meshes, \reviewerOne{and are created using the approach in~\ref{appx_b}}.  The $50\times 50\times 50$ mesh is shown in Figure~\ref{fig:\preftex/mesh}.

\begin{figure}[h]
\centering
\includegraphics[scale=.33,clip=true,trim=0in 0in 1.5in 0.5in]{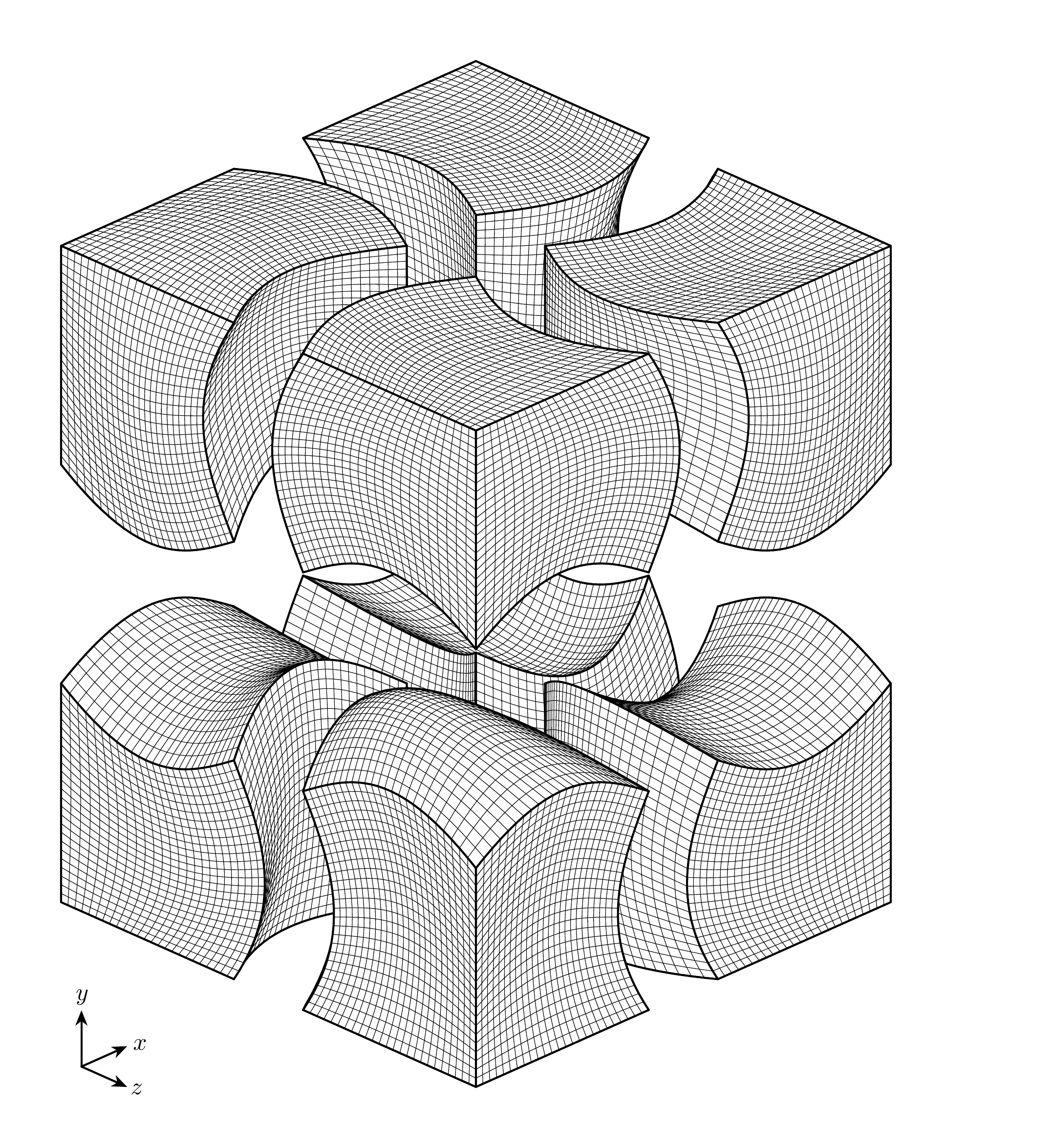}
\caption{\prefix: Mesh with $50\times 50\times 50$ elements.}
\label{fig:\preftex/mesh}
\end{figure}

Table~\ref{tab:\preftex/corr} shows the observed order of accuracy, using the \linfnorm{} of the error, which indicates second-order accuracy $(\globalaccuracy=2)$.  Figure~\ref{fig:\preftex/linf} shows the \linfnorm{} for each of the flow variables.

\begin{table}
\centering
\begin{tabular}{c c c c c c}
\toprule
Mesh & $\rho$ & $u$ & $v$ & $w$ & $T$ \\
\midrule
1--2 & 2.0849 & 1.8731 & 1.9841 & 1.7039 & 1.9404 \\
2--3 & 2.1406 & 1.9923 & 1.9295 & 1.8621 & 1.9774 \\
3--4 & 2.0990 & 2.0115 & 1.9623 & 1.9349 & 1.9922 \\
4--5 & 2.0585 & 2.0100 & 1.9820 & 1.9571 & 1.9964 \\
\bottomrule
\end{tabular}
\caption{\prefix: Observed order of accuracy $\globalaccuracy$ using \linfnorm{} of the error.}
\label{tab:\preftex/corr}
\end{table}

\begin{figure}
\centering
\includegraphics[scale=.64,clip=true,trim=1.2in 0in 1.2in 0in]{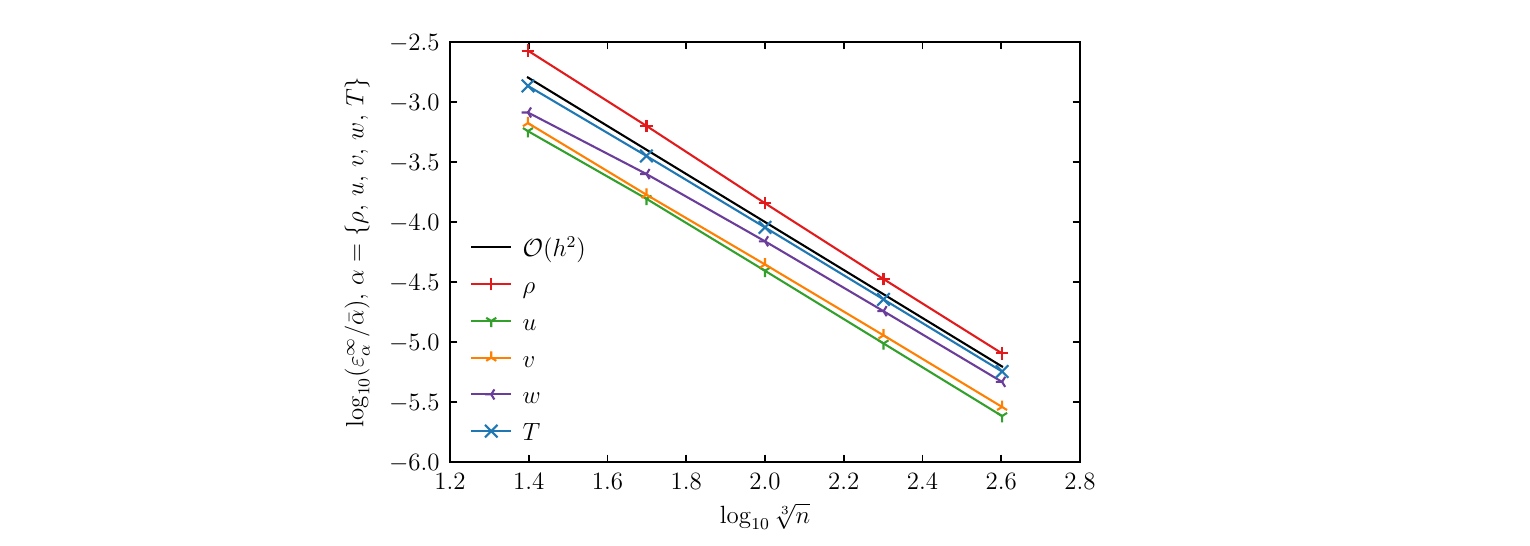}
\caption{\prefix: \linfnorm{} of the error.}
\label{fig:\preftex/linf}
\end{figure}

\FloatBarrier


\subsection{Five-Species Inviscid Flow in Chemical Nonequilibrium} 
This set of tests uses the five-species air model ($\nspecies=5$, $\rhovec=\left\{\rho_{\text{N}_2},\,\rho_{\text{O}_2},\,\rho_\text{NO},\,\rho_\text{N},\,\rho_\text{O}\right\}$) mentioned in Section~\ref{subsec:five_spec}. The flow is inviscid ($\fluxd=0$) and in chemical nonequilibrium ($\wvecdot\ne\mathbf{0}$).  For this type of flow, the reference velocity $\bar{\xvel}$ is determined from a reference Mach number $\bar{\mach}$:
\begin{align}
\bar{\xvel}=\bar{\mach}\sqrt{\gamma \left(\sum_{s=1}^\nspecies \frac{\rho_s}{\rho}\frac{\Runiv}{M_s}\right) \bar{T}}.
\label{eq:ubar5sp}
\end{align}

While these tests do not directly test the thermochemical-source-term implementation (as described in Section~\ref{sec:source_terms}), they do test the coupling of the source term with the differential terms, as well as the spatial discretizations of multiple species and temperatures. 

\subsubsection{2D Supersonic Flow in Thermal Equilibrium using a Manufactured Solution}
\def \prefix {2D MMS, $\nspecies=5$, $T_v=T$, $\wvecdot\ne\mathbf{0}$}
\def \preftex {5sp1T}

For this test, the flow is in thermal equilibrium ($T=T_v$).
The manufactured solutions for this case for $\scalar=\{\rho_{\text{N}_2},\,\rho_{\text{O}_2},\,\rho_{\text{NO}},\,\rho_{\text{N}},\,\rho_{\text{O}},\,\xvel,\,\yvel,\,T\}$ are listed in~\eqref{eq:2d_solutions} and shown in Figure~\ref{fig:2d_solutions}.

The domain is a square with $(x,y)\in[0,\,1]\text{ m}\times [0,\,1]\text{ m}$, and 
$\bar{\rho}_{\text{N}_2}=0.77$ kg/m$^3$,
$\bar{\rho}_{\text{O}_2}=0.20$ kg/m$^3$,
$\bar{\rho}_{\text{NO}}=0.01$ kg/m$^3$,
$\bar{\rho}_{\text{N}}=0.01$ kg/m$^3$,
$\bar{\rho}_{\text{O}}=0.01$ kg/m$^3$,
$\bar{\mach}=2.5$,  
$\bar{\yvel}=\bar{\xvel}$~\eqref{eq:ubar5sp},
$\bar{T}=3500$ K
and 
$\perturbation=0.05$.

Upon solving~\eqref{eq:mms}, the observed order of accuracy $\globalaccuracy$ is computed from \eqref{eq:accuracy_norm} for $\scalar=\{\rho_{\text{N}_2},\,\rho_{\text{O}_2},\,\rho_{\text{NO}},\allowbreak\,\rho_{\text{N}},\allowbreak\,\rho_{\text{O}},\allowbreak\,\xvel,\allowbreak\,\yvel,\,T\}$.
Seven 2D meshes are used, consisting of $25\times 25$, $50\times 50$, $100\times 100$, $200\times 200$, $400\times 400$, $800\times 800$, and $1600\times 1600$ elements.  These meshes are the same nonuniform meshes as those used in Section~\ref{subsubsec:twodinviscidmms}.

Table~\ref{tab:\preftex/corr} shows the observed order of accuracy, using the \linfnorm{} of the error, which indicates second-order accuracy $(\globalaccuracy=2)$.
Figure~\ref{fig:\preftex/linf} shows the \linfnorm{} for each of the flow variables.

\begin{table}
\centering
\begin{tabular}{c c c c c c c c c}
\toprule
Mesh & $\rho_{\text{N}_2}$ & $\rho_{\text{O}_2}$ & $\rho_{\text{NO}}$ & $\rho_{\text{N}}$ & $\rho_{\text{O}}$ & $u$ & $v$ & $T$ \\
\midrule
1--2 & 2.0608 & 2.1382 & 2.0698 & 2.0644 & 2.1885 & 1.8425 & 1.8289 & 1.7351 \\
2--3 & 2.1161 & 2.1219 & 2.1127 & 2.1072 & 2.1697 & 1.8875 & 1.9220 & 1.7923 \\
3--4 & 2.0798 & 2.0813 & 1.8555 & 2.0754 & 2.0971 & 1.9200 & 1.9686 & 1.8525 \\
4--5 & 2.0456 & 2.0458 & 1.8917 & 2.0428 & 2.0806 & 1.9522 & 1.9871 & 1.9079 \\
5--6 & 2.0243 & 2.0243 & 1.9427 & 2.0228 & 2.0529 & 1.9735 & 1.9939 & 1.9485 \\
6--7 & 2.0125 & 2.0125 & 1.9790 & 2.0118 & 2.0318 & 1.9865 & 1.9969 & 1.9737 \\
\bottomrule
\end{tabular}
\caption{\prefix: Observed order of accuracy $\globalaccuracy$ using \linfnorm{} of the error.}
\label{tab:\preftex/corr}
\end{table}

\begin{figure}[!b]
\centering
\includegraphics[scale=.64,clip=true,trim=1.2in 0in 1.2in 0in]{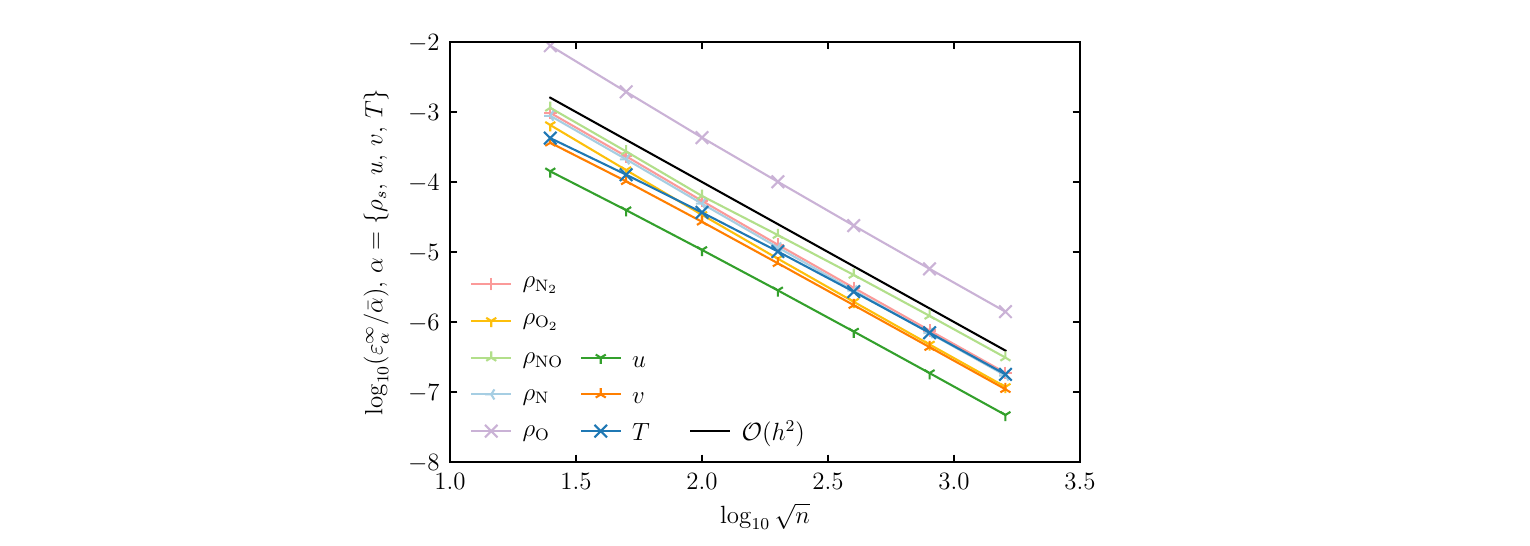}
\caption{\prefix: \linfnorm{} of the error.}
\label{fig:\preftex/linf}
\end{figure}

\FloatBarrier
\subsubsection{2D Hypersonic Flow in Thermal Nonequilibrium using a Manufactured Solution}
\def \prefix {2D MMS, $\nspecies=5$, $T_v\ne T$, $\wvecdot\ne\mathbf{0}$}
\def \preftex {5sp2T}

For this test, the flow is in thermal nonequilibrium ($T\ne T_v$).
The manufactured solutions for this case for $\scalar=\{\rho_{\text{N}_2},\,\rho_{\text{O}_2},\,\rho_{\text{NO}},\,\rho_{\text{N}},\,\rho_{\text{O}},\,\xvel,\,\yvel,\,T,\,T_v\}$ are listed in~\eqref{eq:2d_solutions} and shown in Figure~\ref{fig:2d_solutions}.

The domain is a square with $(x,y)\in[0,\,1]\text{ m}\times [0,\,1]\text{ m}$, and 
$\bar{\rho}_{\text{N}_2}=0.0077$ kg/m$^3$,
$\bar{\rho}_{\text{O}_2}=0.0020$ kg/m$^3$,
$\bar{\rho}_{\text{NO}}=0.0001$ kg/m$^3$,
$\bar{\rho}_{\text{N}}=0.0001$ kg/m$^3$,
$\bar{\rho}_{\text{O}}=0.0001$ kg/m$^3$,
$\bar{\mach}=8$,    
$\bar{\yvel}=\bar{\xvel}$~\eqref{eq:ubar5sp},
$\bar{T}=5000$ K,  
$\bar{T}_v=1000$ K, 
and 
$\perturbation=0.05$.

Upon solving~\eqref{eq:mms}, the observed order of accuracy $\globalaccuracy$ is computed from \eqref{eq:accuracy_norm} for $\scalar=\{\rho_{\text{N}_2},\,\rho_{\text{O}_2},\,\rho_{\text{NO}},\allowbreak\,\rho_{\text{N}},\allowbreak\,\rho_{\text{O}},\allowbreak\,\xvel,\allowbreak\,\yvel,\allowbreak\,T,\,T_v\}$.
Seven 2D meshes are used, consisting of $25\times 25$, $50\times 50$, $100\times 100$, $200\times 200$, $400\times 400$, $800\times 800$, and $1600\times 1600$ elements.  These meshes are the same nonuniform meshes as those used in Section~\ref{subsubsec:twodinviscidmms}.

Table~\ref{tab:\preftex/corr} shows the observed order of accuracy, using the \linfnorm{} of the error, which indicates second-order accuracy $(\globalaccuracy=2)$.
Figure~\ref{fig:\preftex/linf} shows the \linfnorm{} for each of the flow variables.

\begin{table}[!h]
\centering
\begin{tabular}{c c c c c c c c c c}
\toprule
Mesh & $\rho_{\text{N}_2}$ & $\rho_{\text{O}_2}$ & $\rho_{\text{NO}}$ & $\rho_{\text{N}}$ & $\rho_{\text{O}}$ & $u$ & $v$ & $T$ & $T_v$ \\
\midrule
1--2 & 1.5659 & 1.6370 & 1.6555 & 1.6046 & 1.5869 & 1.7742 & 1.7337 & 1.7814 & 1.5545 \\
2--3 & 1.9067 & 1.6944 & 1.6986 & 1.7598 & 1.8819 & 1.8916 & 1.8701 & 1.8768 & 1.9150 \\
3--4 & 1.9868 & 2.0475 & 2.0698 & 2.0477 & 2.0110 & 1.9488 & 1.9357 & 1.9349 & 2.0082 \\
4--5 & 2.0074 & 1.9941 & 2.0138 & 1.9936 & 2.0089 & 1.9752 & 1.9684 & 1.9672 & 2.0168 \\
5--6 & 2.0062 & 1.9939 & 2.0004 & 1.9935 & 2.0061 & 1.9879 & 1.9843 & 1.9836 & 2.0111 \\
6--7 & 2.0037 & 1.9965 & 1.9994 & 1.9962 & 1.9955 & 1.9940 & 1.9922 & 1.9918 & 2.0063 \\
\bottomrule
\end{tabular}
\caption{\prefix: Observed order of accuracy $\globalaccuracy$ using \linfnorm{} of the error.}
\label{tab:\preftex/corr}
\end{table}

\begin{figure}[!h]
\centering
\includegraphics[scale=.64,clip=true,trim=1.2in 0in 1.2in 0in]{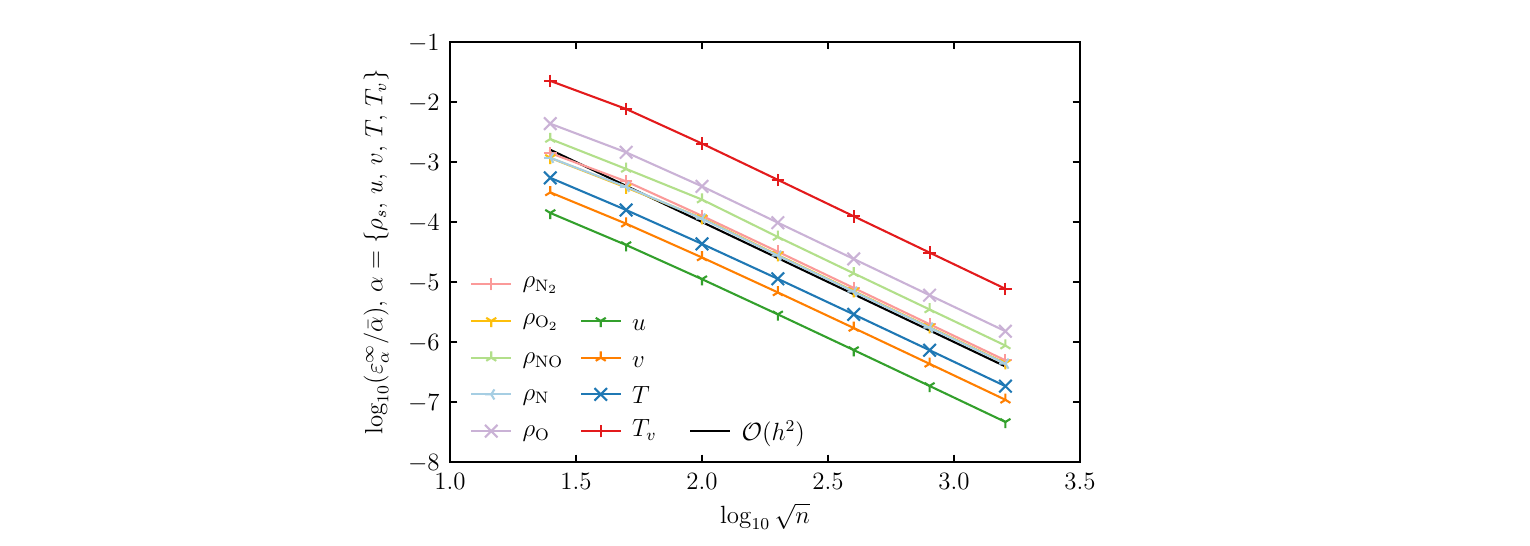}
\caption{\prefix: \linfnorm{} of the error.}
\label{fig:\preftex/linf}
\end{figure}

\FloatBarrier

\section{Verification Techniques for Thermochemical Source Term} 
\label{sec:source_terms}

While measuring the spatial order of accuracy is an effective technique for assessing the discretization, it does not directly reveal errors in the coding of the source term~$\source\left(\state\right)$ in \eqref{eq:conservation}. 
With manufactured solutions, for example, terms containing derivatives in~\eqref{eq:conservation} are evaluated numerically when computing $\residual_h(\state_h)$ in \eqref{eq:mms} and analytically when computing $\residual(\state_\text{MS})$.  
\reviewerOne{The thermochemical source term is} similarly evaluated on both sides of~\eqref{eq:mms}; however, because the algebraic source-term evaluation does not depend on the spatial discretization, these evaluations use the same source code.  Therefore, there are no differences in the evaluation on each side, and, as a result, errors in the source-term implementation are not detected.

To address this limitation, we independently developed a code to compute \reviewerOne{the terms within the source term}, specifically $\Qtv(\rhovec,T,T_v)$, $\evecv(\rhovec,T,T_v)$, and $\wvecdot(\rhovec,T,T_v)$ in \eqref{eq:conservation}.  \reviewerTwo{We compute these terms for many samples of $\{\rhovec,\,T,\,T_v\}$ and compare with those obtained from \sparc{} for a single-cell mesh when initialized to those values with no velocity.  Additionally, we assess the sufficiency of the number of samples by performing convergence studies on distribution properties of the values and on the differences between the two codes.}  

\reviewerTwo{Because this approach only computes the thermochemical source term, it can be considered an extensive unit test.  
Examples of simpler, traditional unit tests include test computations of the forward reaction-rate coefficient $k_{f_r}$ and the equilibrium constant $K_{e_r}$ for each type of reaction, the translational--vibrational energy relaxation time $\tau_{s,m,s'}$ for each collision, and the dependent intermediate computations to obtain $\Qtv$ and $\wvecdot$.  

However, chemical-kinetics models provide individual reaction-rate parameters as a set, with the individual parameters determined in a manner such that a model best matches experimental data in the regime of interest. 
The nonlinear interactions of the individual reactions can lead to much richer model behavior than the simple structure of the reaction-rate forms suggests. 
In addition to detecting errors, our sampling approach attempts to confirm that the performance of the chemical-kinetics model is not sensitive to the particular implementation. 
These remarks apply to the vibrational nonequilibrium model as well, but to a lesser extent, as the models are less complicated and model parameters are more accessible to theoretical determination. 

We have opted for sampling as an expedient approach that can be easily applied to other algebraic terms.  However, it may be more efficient to analyze a chemical-kinetics model in detail to identify regions or surfaces of high sensitivity, and use an adaptive sampling or optimization approach to choose the locations to compare alternative implementations.  Nonetheless, through our approach, we are able to quickly detect errors.}

While it may be instinctive to dismiss these techniques as a typically low-rigor code-to-code comparison, we clarify the distinctive and rigorous features.  
\begin{enumerate}
\item This code is independently developed, using the same models and material properties expected to be employed by \sparc{} but taken directly from the original references.  Alternatively, external software could be used, but, given the variety of published models and material properties, quantifying the agreement and, consequently, assessing the implementation becomes non-trivial.  

\item Because the models and material properties are the same, when computing the difference in the source terms, the required tolerance is tightened from what may typically be a few percent to near machine precision.

\item \reviewerOne{For numerical solutions to partial differential equations, code-to-code comparisons are typically employed for a few canonical cases, through which it is difficult to identify and isolate errors in the numerical-method implementation or attribute differences to specific sources.  On the other hand, our code-to-code comparison targets the portion of the code that manufactured solutions do not assess, and heavily queries conditions covering the thermochemical model's domain of validity.}

\end{enumerate}
The effectiveness of these techniques is demonstrated in Section~\ref{sec:rg_results}.
\section{Thermochemical-Source-Term Verification Results} 
\label{sec:rg_results}

\def \prefix {$\source\left(\state\right)$, $\nspecies=5$, $T_v\ne T$, $\wvecdot\ne\mathbf{0}$}

\reviewerTwo{To assess the correctness of the thermochemical-source-term implementation, we generate $n_\mathcal{S}$ Latin hypercube samples using the ranges and spacings listed in Table~\ref{tab:lhs_samples}, with $n_\mathcal{S}=2^i$, for $i=0,\hdots,17$.  At these samples, we query \sparc{} and the independent code described in Section~\ref{sec:source_terms} to compute $\Qtv(\rhovec,T,T_v)$, $\evecv(\rhovec,T,T_v)$, and $\wvecdot(\rhovec,T,T_v)$ in \eqref{eq:conservation}.}

\begin{table}
\centering
\begin{tabular}{c c c c c}
\toprule
Variable                                    & Minimum   & Maximum  & Units    & Spacing     \\ \midrule
$_{\phantom{\text{N}_2}}\rho_{\text{N}_2} $ & $10^{-6}$ & $10^{1}$ & kg/m$^3$ & Logarithmic \\
$_{\phantom{\text{O}_2}}\rho_{\text{O}_2} $ & $10^{-6}$ & $10^{1}$ & kg/m$^3$ & Logarithmic \\
$_{\phantom{\text{NO} }}\rho_{\text{NO}}  $ & $10^{-6}$ & $10^{1}$ & kg/m$^3$ & Logarithmic \\
$_{\phantom{\text{N}  }}\rho_{\text{N}}   $ & $10^{-6}$ & $10^{1}$ & kg/m$^3$ & Logarithmic \\
$_{\phantom{\text{O}  }}\rho_{\text{O}}   $ & $10^{-6}$ & $10^{1}$ & kg/m$^3$ & Logarithmic \\
$_{\phantom{v}}T_{\phantom{v}}            $ & 100       & 15,000   & K        & Linear      \\
$_{\phantom{v}}T_{v}                      $ & 100       & 15,000   & K        & Linear      \\
\bottomrule
\end{tabular}
\caption{\prefix: Ranges and spacings for Latin hypercube samples of $\rhovec$, $T$, and $T_v$.}
\label{tab:lhs_samples}
\end{table}

\begin{figure}[!t]
\centering%
\begin{subfigure}[t]{0.49\columnwidth}\includegraphics[scale=.64,clip=true,trim=2.2in 0.05in  2.8in 0.15in]{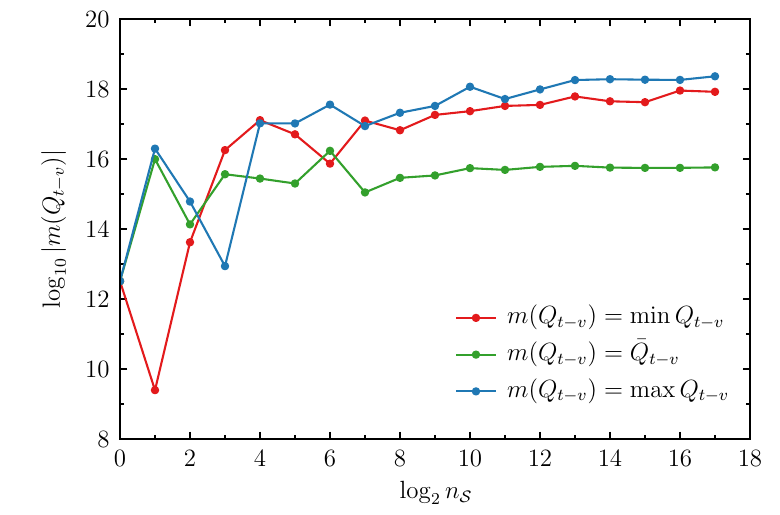}\caption{Minimum, mean, and maximum}\label{fig:Q_tv_range}\end{subfigure} \hfill
\begin{subfigure}[t]{0.49\columnwidth}\includegraphics[scale=.64,clip=true,trim=2.2in 0.05in  2.8in 0.15in]  {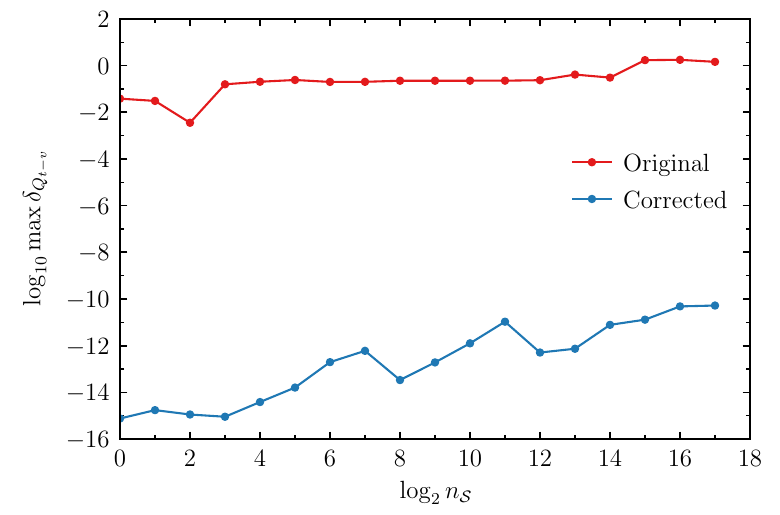}\caption{Maximum relative difference}\label{fig:Q_tv_max_diff}\end{subfigure}
\caption{\reviewerTwo{\prefix: Convergence history of $\Qtv$ sampling.}}
\label{fig:Q_tv_convergence}
\end{figure}

\begin{figure}
\centering%
\begin{subfigure}[t]{0.49\columnwidth}\includegraphics[scale=.64,clip=true,trim=2.2in 0.05in  2.8in 0.15in]{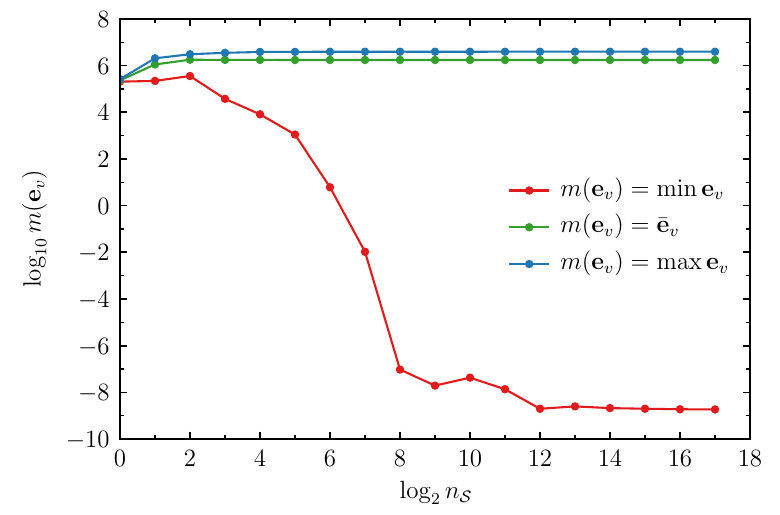}\caption{Minimum, mean, and maximum}\label{fig:e_v_s_range}\end{subfigure} \hfill
\begin{subfigure}[t]{0.49\columnwidth}\includegraphics[scale=.64,clip=true,trim=2.2in 0.05in  2.8in 0.15in]  {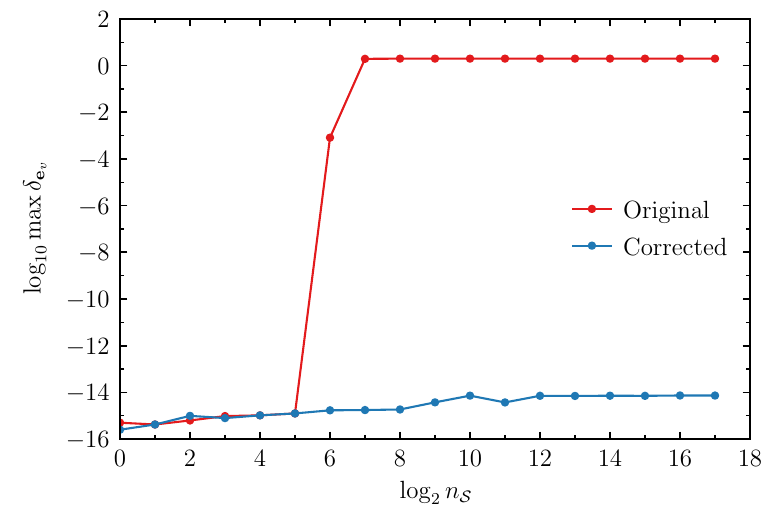}\caption{Maximum relative difference}\label{fig:e_v_s_max_diff}\end{subfigure}
\caption{\reviewerTwo{\prefix: Convergence history of $\evecv$ sampling.}}
\label{fig:e_v_s_convergence}
\end{figure}

\begin{figure}[!b]
\centering%
\begin{subfigure}[t]{0.49\columnwidth}\includegraphics[scale=.64,clip=true,trim=2.2in 0.05in  2.8in 0.15in]{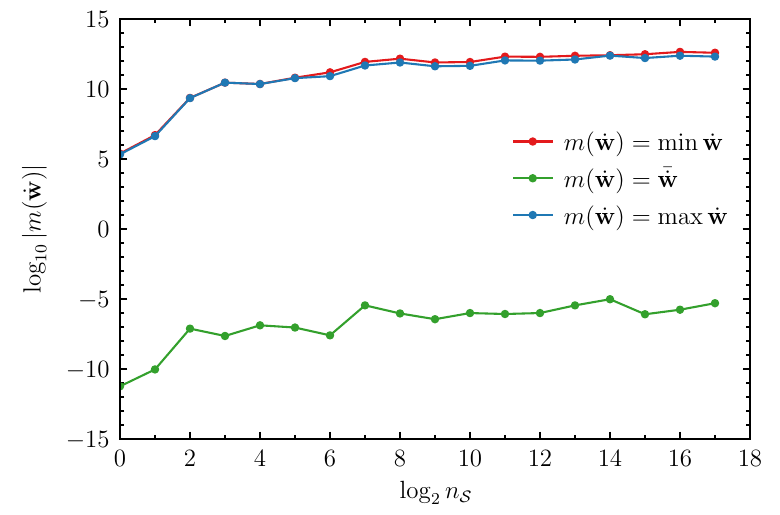}\caption{Minimum, mean, and maximum}\label{fig:w_s_range}\end{subfigure} \hfill
\begin{subfigure}[t]{0.49\columnwidth}\includegraphics[scale=.64,clip=true,trim=2.2in 0.05in  2.8in 0.15in]  {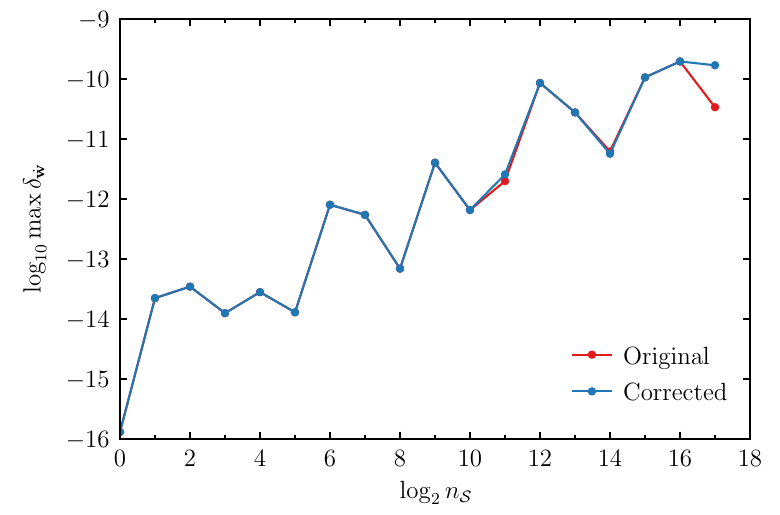}\caption{Maximum relative difference}\label{fig:w_s_max_diff}\end{subfigure}
\caption{\reviewerTwo{\prefix: Convergence history of $\wvecdot$ sampling.}}
\label{fig:w_s_convergence}
\end{figure}

\begin{figure}[!t]
\centering
\includegraphics[scale=.64,clip=true,trim=0in 0.05in  0in 0.15in]{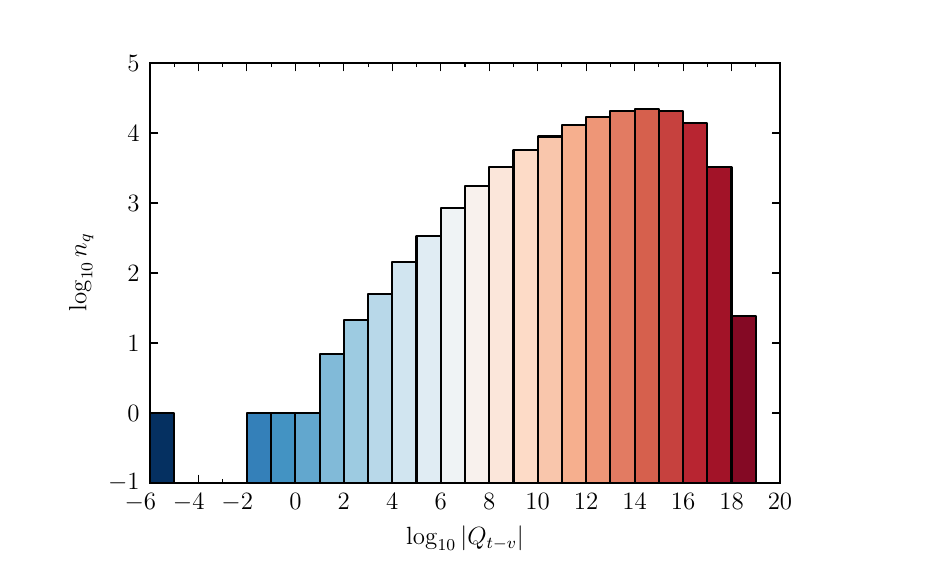}
\caption{\prefix: Absolute values of the translational--vibrational energy exchange $\Qtv$ for $n_\mathcal{S}=2^{17}$.}
\label{fig:Q_tv}
\end{figure}

\begin{figure}
\centering
\includegraphics[scale=.64,clip=true,trim=0in 0.05in  0in 0.15in]{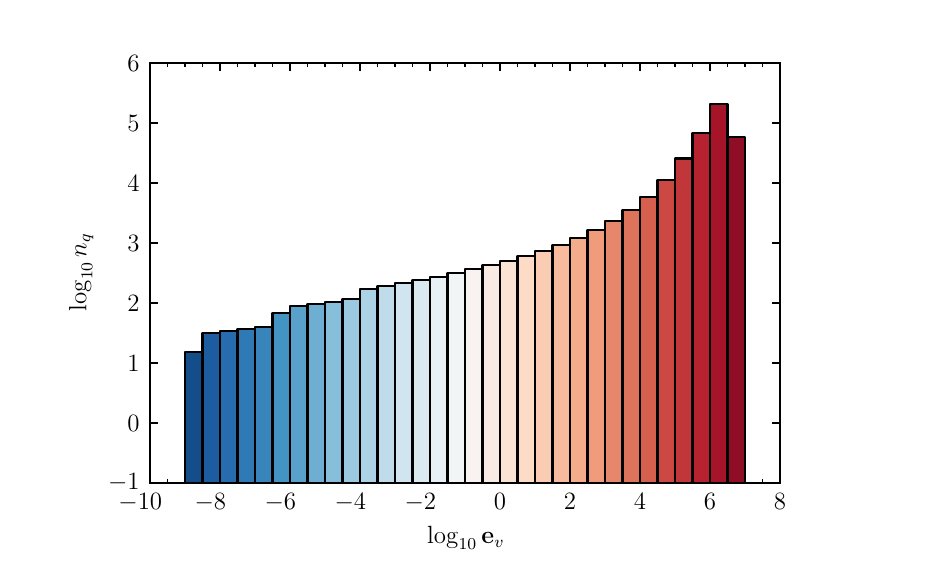}
\caption{\prefix: Vibrational energies per mass $\evecv$ for $n_\mathcal{S}=2^{17}$.}
\label{fig:e_v_s}
\end{figure}

\begin{figure}[!b]
\centering
\includegraphics[scale=.64,clip=true,trim=0in 0.05in  0in 0.15in]{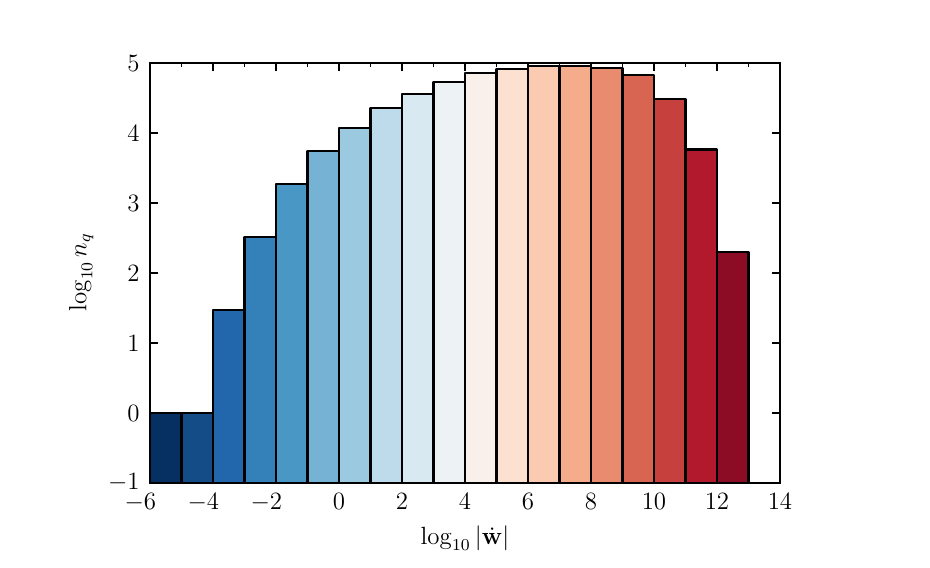}
\caption{\prefix: Absolute values of the mass production rates per volume $\wvecdot$ for $n_\mathcal{S}=2^{17}$.}
\label{fig:w_s}
\end{figure}

\reviewerTwo{%
For each $n_\mathcal{S}$, Figures~\ref{fig:Q_tv_range}, \ref{fig:e_v_s_range}, and \ref{fig:w_s_range} show the minimum, mean, and maximum of the translational--vibrational energy exchange $\Qtv$, the vibrational energies per mass $\evecv$, and the mass production rates per volume $\wvecdot$, as computed from the independent code.  For the vector quantities $\evecv$ and $\wvecdot$, the elements are pooled.  As $n_\mathcal{S}$ is increased, these values are expected to converge; however, because the samples at each $n_\mathcal{S}$ are independently determined, the convergence is not monotonic.  Convergence of these values suggests the distribution of the values is sufficiently resolved, and, therefore, the values are sufficiently represented.

For $n_\mathcal{S}=2^{17}={}$131,072, Figures~\ref{fig:Q_tv}--\ref{fig:w_s} show the distributions of $|\Qtv|$, $\evecv$, and $|\wvecdot|$, as computed from the independent code.  $n_q$ denotes the number of queries within the ranges on the abscissa.  Since $\Qtv$ and $\wvecdot$ can be non-positive, their ranges, as well as those of $\evecv$ are listed in Table~\ref{tab:ranges}.  All of these values vary drastically in magnitude, and, with the exception of $\evecv$, in sign.

For every sample, we compute a symmetric relative difference, defined by
\begin{align}
\delta_\beta = 2\frac{|\beta_{\sparc{}}\phantom{|}-\phantom{|}\beta'|}{\left|\beta_{\sparc{}}\right|+\left|\beta'\right|},
\label{eq:rel_diff}
\end{align}}
where 
$\beta=\left\{
\Qtv ,\,
\evs{\text{N}_2}    ,\,
\evs{\text{O}_2}    ,\,
\evs{\text{NO}}     ,\,
\wdot_{\text{N}_2}  ,\,
\wdot_{\text{O}_2}  ,\,
\wdot_{\text{NO}}   ,\,
\wdot_{\text{N}}    ,\,
\wdot_{\text{O}}    
\right\}$, and the prime denotes computation by the independent code.

\begin{figure}[!t]
\centering
\begin{subfigure}[b]{0.5\textwidth}\centering\includegraphics[scale=.64,clip=true,trim=0.4in 0.05in  0.85in 0.25in]{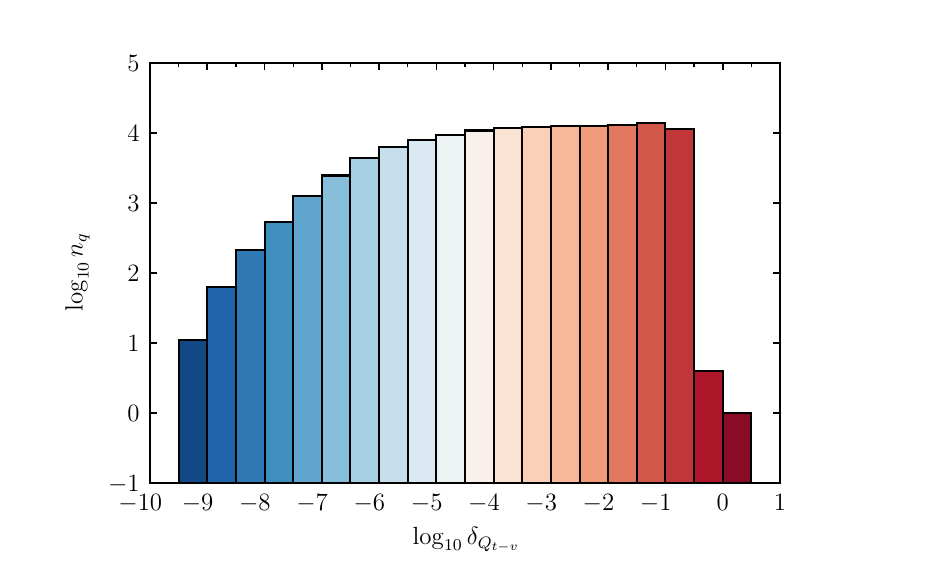}\caption{Original}\label{fig:Q_tv_diff_orig}\end{subfigure}%
\begin{subfigure}[b]{0.5\textwidth}\centering\includegraphics[scale=.64,clip=true,trim=0.4in 0.05in  0.85in 0.25in]  {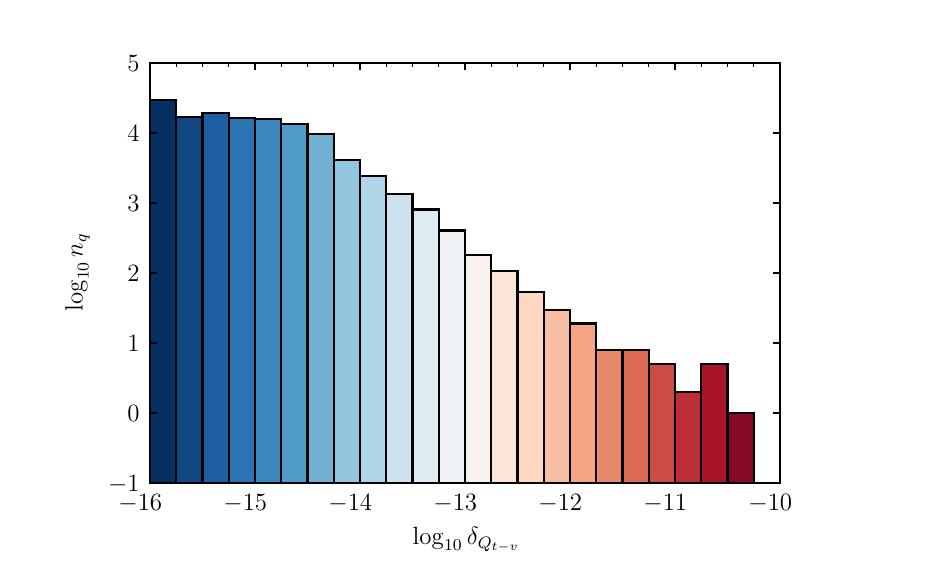}\caption{Corrected}\label{fig:Q_tv_diff_corr}\end{subfigure}%
\caption{\prefix: Relative differences in the translational--vibrational energy exchange $\Qtv$ for $n_\mathcal{S}=2^{17}$, using the original lookup table and convergence criteria (\subref{fig:Q_tv_diff_orig}), and the corrected lookup table with tighter convergence criteria (\subref{fig:Q_tv_diff_corr}).  Queries with $\delta_{\Qtv}=0$ are placed in the lowest bin.}
\label{fig:Q_tv_difference}
\end{figure}

\begin{figure}
\centering
\begin{subfigure}[b]{0.5\textwidth}\centering\includegraphics[scale=.64,clip=true,trim=0.4in 0.05in  0.85in 0.25in]{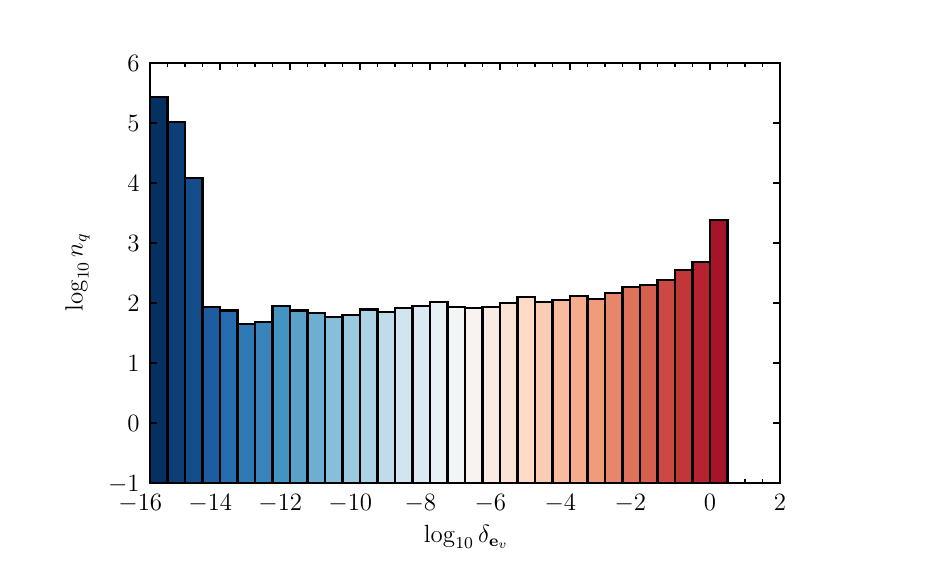}\caption{Original}\label{fig:e_v_s_diff_orig}\end{subfigure}%
\begin{subfigure}[b]{0.5\textwidth}\centering\includegraphics[scale=.64,clip=true,trim=0.4in 0.05in  0.85in 0.25in]  {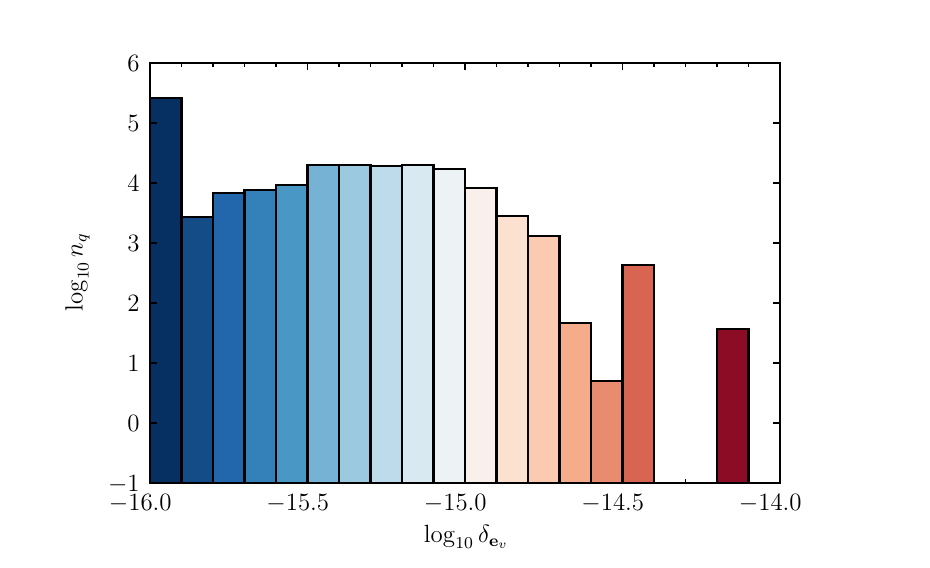}\caption{Corrected}\label{fig:e_v_s_diff_corr}\end{subfigure}%
\caption{\prefix: Relative differences in the vibrational energies per mass $\evecv$ for $n_\mathcal{S}=2^{17}$, using the original convergence criteria (\subref{fig:e_v_s_diff_orig}), and the tighter convergence criteria (\subref{fig:e_v_s_diff_corr}).  Queries with $\delta_{\evecv}=0$ are placed in the lowest bin.}
\label{fig:e_v_s_difference}
\end{figure}

\begin{figure}[!b]
\centering
\begin{subfigure}[b]{0.5\textwidth}\centering\includegraphics[scale=.64,clip=true,trim=0.4in 0.05in  0.85in 0.25in]{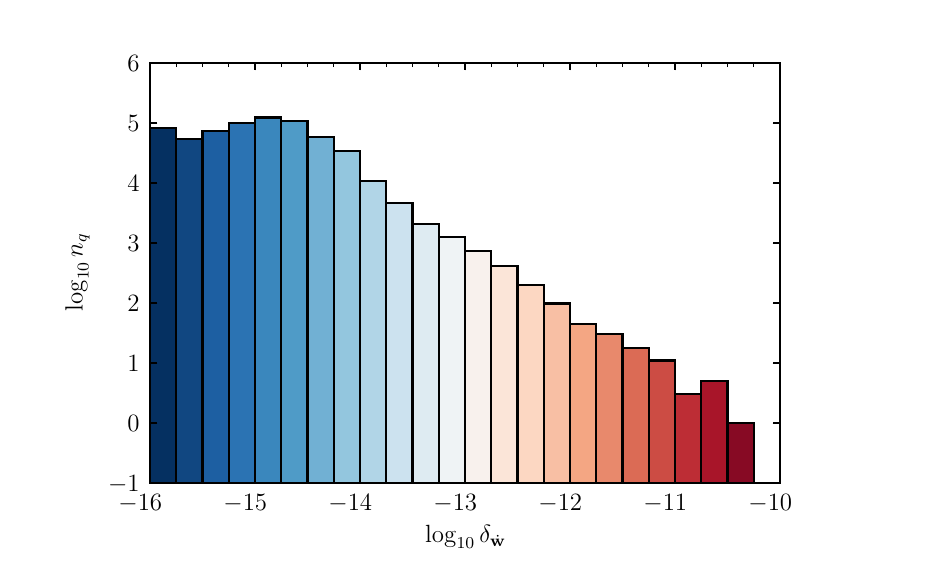}\caption{Original}\label{fig:w_s_diff_orig}\end{subfigure}%
\begin{subfigure}[b]{0.5\textwidth}\centering\includegraphics[scale=.64,clip=true,trim=0.4in 0.05in  0.85in 0.25in]  {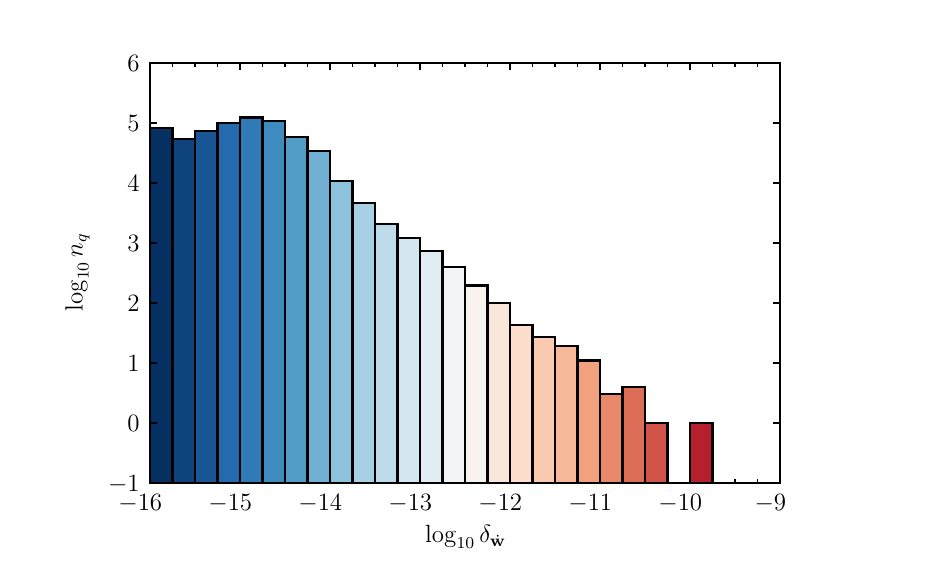}\caption{Corrected}\label{fig:w_s_diff_corr}\end{subfigure}%
\caption{\prefix: Relative differences in the mass production rates per volume $\wvecdot$ for $n_\mathcal{S}=2^{17}$, using the original convergence criteria (\subref{fig:w_s_diff_orig}), and the tighter convergence criteria (\subref{fig:w_s_diff_corr}).  Queries with $\delta_{\wvecdot}=0$ are placed in the lowest bin.}
\label{fig:w_s_difference}
\end{figure}

\reviewerTwo{%
For each $n_\mathcal{S}$, Figures~\ref{fig:Q_tv_max_diff}, \ref{fig:e_v_s_max_diff}, and \ref{fig:w_s_max_diff} show the maximum relative difference across the samples.  As with Figures~\ref{fig:Q_tv_range}, \ref{fig:e_v_s_range}, and \ref{fig:w_s_range}, the maximum relative difference is expected to increase and converge, through not monotonically.  Additionally, for $n_\mathcal{S}=2^{17}$, Figures~\ref{fig:Q_tv_diff_orig}, \ref{fig:e_v_s_diff_orig}, and \ref{fig:w_s_diff_orig} show the relative differences in $\Qtv$, $\evecv$, and $\wvecdot$.
}

As mentioned in Section~\ref{sec:source_terms}, \reviewerTwo{the relative differences~\eqref{eq:rel_diff} are expected to be near machine precision; however, this is clearly not the case for the red curves in Figures~\ref{fig:Q_tv_max_diff} and \ref{fig:e_v_s_max_diff} or the histograms in Figures~\ref{fig:Q_tv_diff_orig} and \ref{fig:e_v_s_diff_orig}}.  As shown in Figure~\ref{fig:Q_tv_diff_orig}, for approximately \reviewerTwo{8.7\%} of the \reviewerTwo{queries}, $\delta_{\Qtv}$ is greater than 10\%, and, for 29\% of the \reviewerTwo{queries}, $\delta_{\Qtv}$ is greater than 1\%.  Additionally, as shown in Figure~\ref{fig:e_v_s_diff_orig}, although $\delta_{\evecv}$ is less than $10^{-12}$ for 99\% of the \reviewerTwo{queries}, $\delta_{\evecv}$ is greater than 100\% for a few of the \reviewerTwo{queries}.  \reviewerTwo{Even with $n_\mathcal{S}=1$, the red curve in Figure~\ref{fig:Q_tv_max_diff} indicates $\delta_{\Qtv}=3.9\%$.  On the other hand, in Figure~\ref{fig:e_v_s_max_diff}, $\delta_{\evecv}$ is within machine precision through $n_\mathcal{S}=32$, but increases by orders of magnitude at $n_\mathcal{S}=64$ and $n_\mathcal{S}=128$.  These observations demonstrate how thirty-two samples are not enough to represent a seven-dimensional space.}

These high relative differences were due to two causes.
\begin{enumerate}
\item The lookup table used by \sparc{} contained incorrect values for the vibrational constants used in~\eqref{eq:tau} for $\text{N}_2$ and $\text{O}_2$ when the colliding species is NO.  These incorrect values introduced an error in $\Qtv$ for all \reviewerTwo{samples}.
\item The convergence criteria specified in the implementation of Newton's method used to compute $T_v$ from $\rho \ev$ was loose.  Though sufficient for most values of $T_v$, these criteria prove unsuitable for low values.  These criteria introduced errors in $\Qtv$ and $\evecv$ for a few \reviewerTwo{samples.  $T_v$ also appears in $T_c$ for dissociative reactions, but the actual impact on $\wvecdot$ was quite small.}  For a converged, steady problem, however, the original convergence criteria is not expected to affect the final solution.
\end{enumerate} 

Upon correcting the lookup-table values and tightening the convergence criteria, we reran the \sparc{} simulations and recomputed \reviewerTwo{$\delta_{\Qtv}$, $\delta_{\evecv}$, and $\delta_{\wvecdot}$, which are shown in Figures~\ref{fig:Q_tv_diff_corr}, \ref{fig:e_v_s_diff_corr}, and \ref{fig:w_s_diff_corr} for $n_\mathcal{S}=2^{17}$.  These results are consistent with our expectations, as all $\delta_{\Qtv}$ values are less than $10^{-10}$; all $\delta_{\evecv}$ values are less than $10^{-14}$; and, with the exception of one query, which is slightly greater, all $\delta_{\wvecdot}$ values are less than $10^{-10}$.

Of the 131,072 $\delta_{\Qtv}$ values, the forty-eight greater than $10^{-12}$ occur when $T$ and $T_v$ have a relative difference of less than 0.2\%.}  As a result, in the numerator of~\eqref{eq:landau_teller}, $\evsm(T)$ and $\evsm(T_v)$ share many of the leading digits; therefore, precision is lost when computing their difference.

\reviewerTwo{Of the 655,360 $\delta_{\wvecdot}$ values computed from the 131,072 samples, 109 are greater than $10^{-12}$.  These slightly elevated differences are a result of the precision loss that can occur from subtraction in~\eqref{eq:ckm}.

The blue curves in Figures~\ref{fig:Q_tv_max_diff}, \ref{fig:e_v_s_max_diff}, and \ref{fig:w_s_max_diff} show how the maximum relative differences vary with respect to $n_\mathcal{S}$ after correcting the lookup-table values and tightening the convergence criteria.  As we expect, these curves are generally increasing and converging.  The change in $T_c$ impacts $\delta_{\wvecdot}$ only about as much as the precision loss.}

\clearpage

\clearpage

\clearpage

\FloatBarrier

\begin{table}[!t]
\centering\reviewerTwo{
\begin{tabular}{c r @{} l r @{} l r @{/} l}
\toprule
Variable                                        & \multicolumn{2}{c}{Minimum}      & \multicolumn{2}{c}{Maximum}   & \multicolumn{2}{c}{Units} \\ \midrule
$_{\phantom{t-v}       }Q      _{t-v}         $ & $-8.3102$ & ${}\times 10^{17}  $ & $2.3066$ & ${}\times 10^{18}$ &  J & m$^3/s$              \\
$_{\phantom{v_{\text{N}_2}}}\evs{\text{N}_2}  $ & $ 1.8510$ & ${}\times 10^{-9}  $ & $3.9668$ & ${}\times 10^{6}$  &  J & kg                   \\
$_{\phantom{v_{\text{O}_2}}}\evs{\text{O}_2}  $ & $ 1.1129$ & ${}\times 10^{-4}  $ & $3.6138$ & ${}\times 10^{6}$  &  J & kg                   \\
$_{\phantom{v_{\text{NO}} }}\evs{\text{NO} }  $ & $ 4.6272$ & ${}\times 10^{-7}  $ & $3.7781$ & ${}\times 10^{6}$  &  J & kg                   \\
$_{\phantom{\text{N}_2}}\wdot_{\text{N}_2}    $ & $-2.1148$ & ${}\times 10^{10}  $ & $4.4677$ & ${}\times 10^{11}$ & kg & m$^3/s$              \\
$_{\phantom{\text{O}_2}}\wdot_{\text{O}_2}    $ & $-1.9297$ & ${}\times 10^{12}  $ & $1.1891$ & ${}\times 10^{11}$ & kg & m$^3/s$              \\
$_{\phantom{\text{NO}} }\wdot_{\text{NO}}     $ & $-3.8719$ & ${}\times 10^{12}  $ & $1.8000$ & ${}\times 10^{12}$ & kg & m$^3/s$              \\
$_{\phantom{\text{N}}  }\wdot_{\text{N}}      $ & $-8.4915$ & ${}\times 10^{11}  $ & $1.7631$ & ${}\times 10^{12}$ & kg & m$^3/s$              \\
$_{\phantom{\text{O}}  }\wdot_{\text{O}}      $ & $-3.6139$ & ${}\times 10^{11}  $ & $2.0645$ & ${}\times 10^{12}$ & kg & m$^3/s$              \\
\bottomrule
\end{tabular}}
\caption{\prefix: Ranges for $\Qtv$,  $\evecv$,  and $\wvecdot$ for $n_\mathcal{S}=2^{17}$.}
\phantom{\strut}
\label{tab:ranges}
\end{table}

\reviewerTwo{Though the incorrect lookup-table values introduced high relative differences immediately, the convergence criteria required more samples to detect their looseness.  The convergence of the distribution properties in Figures~\ref{fig:Q_tv_range}, \ref{fig:e_v_s_range}, and \ref{fig:w_s_range} and the convergence and bounds of the errors in Figures~\ref{fig:Q_tv_max_diff}, \ref{fig:e_v_s_max_diff}, and \ref{fig:w_s_max_diff} provide insight into how many samples are enough to assess the correctness of the thermochemical-source-term implementation over the ranges listed in Table~\ref{tab:lhs_samples}.}

\reviewerOne{To determine the impact of these modifications, we reran a high-enthalpy (20 MJ/kg), hypersonic, laminar double-cone flow case after correcting the lookup table values and tightening the convergence criteria.  Relative to the original values and convergence criteria, we observed up to a 1.4\% and 2.7\% change in the pressure and heat flux, respectively, on the surface of the body.}


\section{Conclusions} 
\label{sec:conclusions}

In this paper, we presented our code-verification techniques for hypersonic reacting flows in thermochemical nonequilibrium.  
\reviewerOne{%
To assess the spatial accuracy, we employed manufactured and exact solutions with the \linfnorm{} and \lonenorm{}.  These approaches revealed the impact of the lower-order boundary conditions.
To assess the algebraic thermochemical source term, we queried an independent code, with the expectation that agreement be near machine precision.  We queried the independent code many times \reviewerTwo{and studied the convergence properties}.  This approach revealed the impact of erroneous lookup table entries and insufficiently tight convergence criteria.

}
While the scope of this paper has been limited to flows in vibrational nonequilibrium with five species that undergo dissociation and exchange reactions, these techniques could be analogously extended to address more complex flows in rotational and electronic nonequilibrium that contain additional species capable of undergoing ionization reactions.  \reviewerTwo{However, it is important to perform a convergence study to determine whether the number of samples sufficiently spans the ranges of the dependencies.}

\section*{Acknowledgments}

The authors thank Derek Dinzl, Travis Fisher, Micah Howard, and Ross Wagnild for their valuable assistance with \sparc{} and the underlying models and properties, as well as Neil Matula for his insightful feedback.
This paper describes objective technical results and analysis. Any subjective views or opinions that might be expressed in the paper do not necessarily represent the views of the U.S. Department of Energy or the United States Government.
Sandia National Laboratories is a multimission laboratory managed and operated by National Technology and Engineering Solutions of Sandia, LLC., a wholly owned subsidiary of Honeywell International, Inc., for the U.S. Department of Energy's National Nuclear Security Administration under contract DE-NA-0003525.
\appendix
\renewcommand*{\thesection}{Appendix \Alph{section}}
\setcounter{table}{0}
\renewcommand{\thetable}{A\arabic{table}}
\clearpage
\section{Five-Species Air Model Properties} 
\label{appx}

The molecular weights $M_s$ and formation enthalpies $h_s^o$ for the five species are tabulated in Table~\ref{tab:species}.  

\begin{table}[htb!]
\centering
\begin{tabular}{c c c c}
\toprule
$s$ & Species & $M_s$ [g/mol] & $h_s^o$, 0 K [J/kg]\\
\midrule
1 & $\Ntwo$ & 28.016 & $0\phantom{.000000\times 10^0}$\\
2 & $\Otwo$ & 32.000 & $0\phantom{.000000\times 10^0}$\\
3 & NO      & 30.008 & $2.996123\times 10^6$ \\
4 & N       & 14.008 & $3.362161\times 10^7$ \\
5 & O       & 16.000 & $1.542000\times 10^7$ \\
\bottomrule
\end{tabular}
\caption{Five-species air model: Molecular weights $M_s$ and formation enthalpies $h_s^o$~\cite{candler_1988}.}
\label{tab:species}
\end{table} 
\FloatBarrier

The characteristic vibrational temperatures $\thetavsm$~\cite{candler_1988} and collision-limiting cross sections at 50,000 K $\sigma_{v_s}'$~\cite{park_1993,park_1994} are tabulated in Table~\ref{tab:theta_v}.

\begin{table}[htb!]
\centering
\begin{tabular}{c c c c c}
\toprule
$s$ & Species & $\nvs$ & $\thetavsm$ [K] & $\sigma_{v_s}'$ [m$^2$]\\ \midrule
1   & $\Ntwo$ & 1      & 3395                   & $3\times 10^{-21}$ \\
2   & $\Otwo$ & 1      & 2239                   & $3\times 10^{-21}$ \\
3   & NO      & 1      & 2817                   & $3\times 10^{-21}$ \\
\bottomrule
\end{tabular}
\caption{Five-species air model: Characteristic vibrational temperatures $\thetavsm$~\cite{candler_1988} and collision-limiting cross sections at 50,000 K $\sigma_{v_s}'$~\cite{park_1993,park_1994}.}
\label{tab:theta_v}
\end{table}
\FloatBarrier

The vibrational constants $a_{s,m,s'}$ and $b_{s,m,s'}$ for Equation~\eqref{eq:tau}, associated with the reactions listed in Table~\ref{tab:reactions}, are listed in Table~\ref{tab:ab}. 

\begin{table}[htb!]
\centering
\begin{tabular}{c c c c c c c}
\toprule
            & \multicolumn{2}{c}{$\Ntwo$}                & \multicolumn{2}{c}{$\Otwo$}                & \multicolumn{2}{c}{NO}      \\
              \cmidrule(lr){2-3}                           \cmidrule(lr){4-5}                           \cmidrule(lr){6-7}
$s'       $ & $a_{s,m,s'}$ & $b_{s,m,s'}$                & $a_{s,m,s'}$ & $b_{s,m,s'}$                & $a_{s,m,s'}$ & $b_{s,m,s'}$ \\ \midrule
$\Ntwo    $ & \multicolumn{2}{c}{Equation~\eqref{eq:ab}} & \multicolumn{2}{c}{Equation~\eqref{eq:ab}} & 49.5         & 0.0420       \\
$\Otwo    $ & \multicolumn{2}{c}{Equation~\eqref{eq:ab}} & \multicolumn{2}{c}{Equation~\eqref{eq:ab}} & 49.5         & 0.0420       \\
$\text{NO}$ & \multicolumn{2}{c}{Equation~\eqref{eq:ab}} & \multicolumn{2}{c}{Equation~\eqref{eq:ab}} & 49.5         & 0.0420       \\
$\text{N} $ & \multicolumn{2}{c}{Equation~\eqref{eq:ab}} & 72.4         & 0.0150                      & 49.5         & 0.0420       \\
$\text{O} $ & 72.4         & 0.0150                      & 47.7         & 0.0590                      & 49.5         & 0.0420       \\
\bottomrule
\end{tabular}
\caption{Five-species air model: Vibrational constants $a_{s,m,s'}$ and $b_{s,m,s'}$~\cite{park_1993}.}
\label{tab:ab}
\end{table}
\FloatBarrier
\clearpage
The stoichiometric coefficients $\alpha_{s,r}$ and $\beta_{s,r}$ associated with the reactions listed in Table~\ref{tab:reactions} are listed in Table~\ref{tab:stoichiometric_coefficients}.  

\begin{table}[htb!]
\centering
\begin{tabular}{c c c c c c c c c c c}
\toprule
    & \multicolumn{5}{c}{$\alpha_{s,r}$} & \multicolumn{5}{c}{$\beta_{s,r}$} \\
     \cmidrule(lr){2-6}                \cmidrule(lr){7-11}
$r$ & $\Ntwo$ & $\Otwo$ & NO & N & O & $\Ntwo$ & $\Otwo$ & NO & N & O \\ \midrule
\pz1   & 2    &         &    &   &   & 1       &         &    & 2 &   \\
\pz2   & 1    & 1       &    &   &   &         & 1       &    & 2 &   \\
\pz3   & 1    &         & 1  &   &   &         &         & 1  & 2 &   \\
\pz4   & 1    &         &    & 1 &   &         &         &    & 3 &   \\
\pz5   & 1    &         &    &   & 1 &         &         &    & 2 & 1 \\
\pz6   & 1    & 1       &    &   &   & 1       &         &    &   & 2 \\
\pz7   &      & 2       &    &   &   &         & 1       &    &   & 2 \\
\pz8   &      & 1       & 1  &   &   &         &         & 1  &   & 2 \\
\pz9   &      & 1       &    & 1 &   &         &         &    & 1 & 2 \\
  10   &      & 1       &    &   & 1 &         &         &    &   & 3 \\
  11   & 1    &         & 1  &   &   & 1       &         &    & 1 & 1 \\
  12   &      & 1       & 1  &   &   &         & 1       &    & 1 & 1 \\
  13   &      &         & 2  &   &   &         &         & 1  & 1 & 1 \\
  14   &      &         & 1  & 1 &   &         &         &    & 2 & 1 \\
  15   &      &         & 1  &   & 1 &         &         &    & 1 & 2 \\
  16   & 1    &         &    &   & 1 &         &         & 1  & 1 &   \\
  17   &      &         & 1  &   & 1 &         & 1       &    & 1 &   \\
\bottomrule
\end{tabular}
\caption{Five-species air model: Stoichiometric coefficients $\alpha_{s,r}$ and $\beta_{s,r}$.}
\label{tab:stoichiometric_coefficients}
\end{table}
\FloatBarrier

The reaction-rate-coefficient dependencies $C_{f_r}$, $\eta_r$, and $\theta_r$ associated with the reactions listed in Table~\ref{tab:reactions} are listed in Table~\ref{tab:rrc_coefs}. 

\begin{table}[htb!]
\centering
\begin{tabular}{c c c c}
\toprule
$r$ & $C_{f_r}$ [CGS] & $\eta_r$ & $\theta_r$ [K] \\
\midrule
\pz1   & $7.0\times 10^{21}$ &           $-1.6$ &   113,200 \\
\pz2   & $7.0\times 10^{21}$ &           $-1.6$ &   113,200 \\
\pz3   & $7.0\times 10^{21}$ &           $-1.6$ &   113,200 \\
\pz4   & $3.0\times 10^{22}$ &           $-1.6$ &   113,200 \\
\pz5   & $3.0\times 10^{22}$ &           $-1.6$ &   113,200 \\
\pz6   & $2.0\times 10^{21}$ &           $-1.5$ & \pz59,500 \\
\pz7   & $2.0\times 10^{21}$ &           $-1.5$ & \pz59,500 \\
\pz8   & $2.0\times 10^{21}$ &           $-1.5$ & \pz59,500 \\
\pz9   & $1.0\times 10^{22}$ &           $-1.5$ & \pz59,500 \\
  10   & $1.0\times 10^{22}$ &           $-1.5$ & \pz59,500 \\
  11   & $5.0\times 10^{15}$ & $\phantom{-}0.0$ & \pz75,500 \\
  12   & $5.0\times 10^{15}$ & $\phantom{-}0.0$ & \pz75,500 \\
  13   & $1.1\times 10^{17}$ & $\phantom{-}0.0$ & \pz75,500 \\
  14   & $1.1\times 10^{17}$ & $\phantom{-}0.0$ & \pz75,500 \\
  15   & $1.1\times 10^{17}$ & $\phantom{-}0.0$ & \pz75,500 \\
  16   & $6.4\times 10^{17}$ &           $-1.0$ & \pz38,400 \\
  17   & $8.4\times 10^{12}$ & $\phantom{-}0.0$ & \pz19,400 \\
\bottomrule
\end{tabular}
\caption{Five-species air model: Reaction-rate-coefficient dependencies $C_{f_r}$, $\eta_r$, and $\theta_r$~\cite{park_1990}.}
\label{tab:rrc_coefs}
\end{table}
\FloatBarrier
\clearpage
The equilibrium-constant coefficients, for Equation~\eqref{eq:keq}, associated with the reactions listed in Table~\ref{tab:reactions} are listed in Table~\ref{tab:kcr_coefs}. 

\begin{table}[htb!]
\centering
\begin{tabular}{c c c c c c c}
\toprule
$r$ & $A_{1_r}$ & $A_{2_r}$ & $A_{3_r}$ & $A_{4_r}$ & $A_{5_r}$ \\
\midrule
\pz1--5\pz& $\phantom{-}1.606000$ & $\phantom{-}1.57320$ & $\phantom{-}1.39230$ &   $-11.53300$ &           $-0.0045430$ \\
\pz6--10  & $\phantom{-}0.641830$ & $\phantom{-}2.42530$ & $\phantom{-}1.90260$ & \pz$-6.62770$ & $\phantom{-}0.0351510$ \\
11--15    & $\phantom{-}0.638170$ & $\phantom{-}0.68189$ & $\phantom{-}0.66336$ & \pz$-7.57730$ &           $-0.0110250$ \\
  16      & $\phantom{-}0.967940$ & $\phantom{-}0.89131$ & $\phantom{-}0.72910$ & \pz$-3.95550$ & $\phantom{-}0.0064880$ \\
  17      &           $-0.003732$ &           $-1.74340$ &           $-1.23940$ & \pz$-0.94952$ &           $-0.0461820$ \\
\bottomrule
\end{tabular}
\caption{Five-species air model: Equilibrium-constant coefficients $A_{1_r}$, $A_{2_r}$, $A_{3_r}$, $A_{4_r}$, and $A_{5_r}$~\cite{park_1990}.}
\label{tab:kcr_coefs}
\end{table}
\FloatBarrier
\section{Approach to Creating Nonuniform Meshes} 
\label{appx_b}

\reviewerOne{%
To generate the nonuniform meshes in Figures~\ref{fig:twodinviscidmms/mesh} and~\ref{fig:threedinviscidmms/mesh}, we employ the following nonlinear transformation from $(\xi,\eta,\zeta)\in [0,\,1]\times[0,\,1]\times[0,\,1]$ to $(x,y,z)\in[0,\,L_x]\times[0,\,L_y]\times[0,\,L_z]$:
\begin{align*}
\mathbf{x} = \mathbf{L}\boldsymbol{\xi} + \mathbf{A}\hat{\boldsymbol{\xi}},
\end{align*}
where $\mathbf{x}=\{x,\,y,\,z\}^T$, $\mathbf{L}=\mathrm{diag}\{L_x,\,L_y,\,L_z\}$, $\boldsymbol{\xi}=\{\xi,\,\eta,\,\zeta\}^T$, $\hat{\boldsymbol{\xi}}=\mathbf{L}\left(\tfrac{1}{2}-\boldsymbol{\xi}\right)$, and
\begin{align*}
\mathbf{A} = \left[ 
\begin{matrix}
\alpha\left(s_{\xi,\eta} + s_{\xi,\zeta}\right) &           -\beta s_{\xi,\eta}                    & \phantom{-}\beta s_{\xi,\zeta}
\\[.5em]
\phantom{-}\beta s_{\eta,\xi}                   & \alpha\left(s_{\eta,\xi} + s_{\eta,\zeta}\right) &           -\beta s_{\eta,\zeta}
\\[.5em]
          -\beta s_{\zeta,\xi}                  & \phantom{-}\beta s_{\zeta,\eta}                   &  \alpha\left(s_{\zeta,\xi} + s_{\zeta,\eta}\right) 
\end{matrix}
\right],
\end{align*}
with $\alpha=1-\cos\tfrac{\pi}{6}$, $\beta=\sin\tfrac{\pi}{6}$, and $s_{\xi',\eta'}=\sin(\pi\xi')\sin(\pi\eta')$.  For the mesh in Figure~\ref{fig:twodinviscidmms/mesh}, $\zeta=z=0$.}


\addcontentsline{toc}{section}{\refname}
\bibliographystyle{elsarticle-num}
\bibliography{sparc}

\end{document}